\documentclass[aps,superscriptaddress,nofootinbib]{revtex4}
\usepackage{graphicx,amssymb,amsbsy,bm,amsmath}
\usepackage{epsfig,rotating}
\usepackage{dsfont}

\numberwithin{equation}{section}
\newcommand{\be}{\begin{equation}}
\newcommand{\ee}{\end{equation}}
\newcommand{\bea}{\begin{eqnarray}}
\newcommand{\eea}{\end{eqnarray}}

\newcommand{\no}{\noindent}
\newcommand{\la}{\lambda}
\newcommand{\si}{\sigma}
\newcommand{\vk}{\vec{k}}
\newcommand{\vx}{\vec{x}}
\newcommand{\om}{\omega}
\newcommand{\Om}{\Omega}
\newcommand{\ga}{\gamma}

\newcommand{\app}{\approx}
\newcommand{\uvk}{\widehat{\bf{k}}}
\newcommand{\OM}{\overline{M}}

\begin{document}

\title{Space-time propagation of neutrino wave packets at high temperature and
density}
\author{\bf C. M. Ho}
\email{cmho@phyast.pitt.edu}\affiliation{  Department of Physics and
Astronomy, University of Pittsburgh, Pittsburgh, Pennsylvania 15260,
USA }
\author{\bf D. Boyanovsky}
\email{boyan@pitt.edu}\affiliation{  Department of Physics and
Astronomy, University of Pittsburgh, Pittsburgh, Pennsylvania 15260,
USA }

\date{\today}

\begin{abstract}

We study the space-time evolution of ``flavor'' neutrino
wave-packets at finite temperature and density in the early Universe
prior to BBN. We implement non-equilibrium field theory methods and
linear response to study the space-time evolution directly from the
effective Dirac equation in the medium. There is a rich hierarchy of
time scales associated with transverse and longitudinal dispersion
and coherence. A phenomenon of ``freezing of coherence'' is a result
of a competition between longitudinal dispersion and the separation
of wave-packets of propagating modes in the medium. Near a resonance
the coherence and oscillation time scales are enhanced by a factor
$1/\sin2\theta$ compared to the vacuum. Collisional relaxation via
charged and neutral currents occurs on time scales much shorter than
the coherence time scale and for small vacuum mixing angle, shorter
than the oscillation scale. Assuming that the momentum spread of the
initial wave packet is determined by the large angle scattering mean
free path of charged leptons, we find that the transverse dispersion
time scale is the shortest and is responsible for a large
suppression in both the survival and transition probabilities on
time scales much shorter than the Hubble time. For small mixing
angle the oscillation time scale is \emph{longer} than the
collisional relaxation scale. The method also yields the evolution
of right-handed wave packets. Corrections to the oscillation
frequencies emerge from wave-packet structure as well as from the
energy dependence of mixing angles in the medium.

\end{abstract}


\maketitle

\section{Introduction}

Neutrinos are the bridge between astrophysics, cosmology, particle
physics and nuclear physics \cite{book1,book2,book3,raffelt}. In
recent years, there has been  increasing experimental evidence that
confirm  that   neutrinos are massive and  oscillate between
different flavors\cite{giunti,smirnov, bilenky, haxton, grimus}
providing the first indisputable hint of \emph{new physics} beyond
the Standard Model.

Neutrino mixing and oscillations in extreme conditions of high
temperature and density play a fundamental role in astrophysics and
cosmology\cite{dolgov,pastor,gouvea,prakash,reddy,yakovlev,barko}.
Resonant flavor mixing due to Mikheyev-Smirnov-Wolfenstein (MSW)
effect can provide a concrete explanation to the solar neutrino
problem \cite{MSWI, MSWII}. During  Big Bang Nucleosynthesis (BBN)
 neutrino oscillations may result in corrections to  the abundance of electron
neutrinos\cite{dolgov}. This in turn  changes the neutron-to-proton
ratio, affecting the mass fraction of $^{4}He$ (see ref.
\cite{dolgov} and references therein). Neutrino oscillations violate
lepton number leading to the possibility that the cosmological
baryon asymmetry may originate in the lepton
sector\cite{fukugita,yanagida,buch}.

Neutrino propagation  in a cold medium has been first studied by
Wolfenstein\cite{MSWI} who included the refractive index from
electron neutrinos. Early studies    focused on the neutrino
dispersion relations and damping rates at the temperature limit
relevant for stellar evolution or BBN \cite{notzold}. This work has
been extended to include charged leptons, neutrinos and nucleons in
the thermal medium \cite{dolivoDR}. The matter effects of neutrino
oscillations in the early universe have been investigated in
\cite{barbieri,enqvist,notzold,dolgov}. More Recently, a
non-equilibrium field theoretical description of neutrino
oscillations in the early universe in the real-time formulation has
been reported in \cite{hoboya}.

Kayser\cite{Kayser} first pointed out subtle but important caveats
in the  vacuum oscillation formula obtained from the standard plane
wave treatment, which result from  assuming a definite neutrino
momentum for different mass eigenstates. He showed that
  knowledge of momentum allows experiments to
distinguish different neutrino mass eigenstates,  essentially
destroying the oscillation pattern. He then proposed a wave-packet
treatment of neutrino oscillations, in which the neutrino momentum
is spread out. Since then, the wave-packet approach has been studied
by many authors in both quantum mechanical
\cite{book1,book2,JRich,CG1991,CG1992medium,CG1998,Kiers} and field
theoretical \cite{CG1993,CG2002,Stockinger,Cardall,Beuthe}
frameworks, including the study of oscillations of neutrinos
produced and detected in crystals\cite{crystal}.

The quantum mechanical approach usually refers to the intermediate
wave-packet model in which each propagating mass eigenstate of
neutrino is associated with a wave-packet \cite{CG1991}. This model
eliminates some of the problems in the plane wave treatment although
several conceptual questions remaining unsettled\cite{JRich}. See
\cite{Beuthe} and references therein for detailed descriptions of
these issues. A field theoretical approach is the external
wave-packet model \cite{CG1993} in which the oscillating neutrino is
represented by an internal line of a Feynman diagram, while the
source and the detector are respectively described by in-coming and
out-going wave-packets. A recent review\cite{Beuthe} presents the
different approaches, summarizes their advantages and caveats and
includes the dispersion of wave packets in the study.

An important physical consequence of the wave-packet description of
neutrino evolution is the concept of the   coherence length beyond
which neutrino oscillations vanish. A ``flavor neutrino''
wave-packet is a linear superposition of wave-packets of mass
eigenstates. The different mass states entail a difference in the
group velocity and an eventual separation of the wave-packets
associated with mass eigenstates. This separation results in a
progressive loss of coherence as overlaps between the wave packets
dimishes. See for example\cite{CG1998} for an early explanation. In
an actual source-detector experimental setup, the observation time
is usually not measured and  is commonly integrated out in a
wave-packet treatment\cite{CG1998}. This leads to a localization
term in the vacuum oscillation formula, which states that neutrino
oscillations are suppressed if the spatial uncertainty is much
larger than the oscillation length\cite{CG1998}.

The coherence of neutrino oscillations in matter has been studied
within a geometrical representation in \cite{CG1992medium}, but the
medium oscillation formula was not derived. While most of the
studies focus on reproducing the standard vacuum oscillation
formula, a    consistent study of neutrino mixing and propagation in
a medium \emph{in real time} has not yet emerged.

While in the vacuum the space-time propagation can be studied in the
wave-packet approach by focusing on the space-time evolution of
initially prepared single particle ``flavor states'', the study of
the space-time evolution   in a medium at finite temperature and
density requires a density matrix description.

To the best of our knowledge, a full finite temperature field
theoretical treatment of the space-time propagation of neutrino
wave-packets   in a medium including medium corrections and
dispersion dynamics has not yet been offered. We consider this study
an important aspect of the program to study  the non-equilibrium
evolution of neutrinos in the early Universe. Detailed studies have
shown that neutrino oscillations and self-synchronization lead to
flavor equilibration before BBN\cite{fuller,luna,doleq,wong,aba},
beginning at a temperature of $T \sim 30 MeV$ with complete flavor
equilibration among all flavors at $T \sim 2 MeV$ \cite{doleq}. If
neutrinos are produced in the form of spatially localized
wave-packets rather than extended plane waves before BBN, the two
mass eigenstates separate progressively during propagation  due to
the small difference in group velocities. A significant amount of
neutrino oscillations, which are crucial for ``flavor equalization''
requires that the two mass eigenstates overlap appreciably
throughout their propagation, namely the coherence time scale should
be sufficiently large to ensure that ``flavor equilibration''
through oscillations is effective. Therefore, it is important to
pursue a full field theoretical study of neutrino wave-packet
propagation in the medium directly in \emph{real time} to determine
the relevant time scales for coherence to be maintained and to
identify the processes that contribute to its loss.

\subsection{The main goals of this article}
In this article we study the space-time evolution of neutrino wave
packets in extreme environments at high temperature and density,
conditions that prevail in the early Universe or during supernovae
explosions. Our goals are the following:
\begin{itemize}
\item{i:) to provide a consistent and systematic non-equilibrium
field-theoretical formulation to study the space-time evolution of
initially prepared neutrino wave packets at finite temperature and
density. This goal requires a treatment of the space-time evolution
in terms of the  density matrix, which goes beyond the usual
treatment in terms of single particle states. To achieve this goal
we implement a recently developed method\cite{hoboya} to study
non-equilibrium aspects of neutrino propagation in a medium as an
\emph{initial value problem} in linear response. This method yields
the effective Dirac equation of motion for the expectation value of
the neutrino field induced by an external source. The effective
Dirac equation in the medium includes self-energy contributions from
charged and neutral currents up to one loop. }

\item{ii:) To assess the different time scales associated with
wave-packet dispersion, coherence and oscillations including the
medium effects, in particular near possible resonances in the
in-medium mixing angles. This is achieved by solving the effective
Dirac equation in the medium, which includes self-energy
corrections, as an initial value problem. The initial  wave-packet
configuration is ``prepared'' by an external source term in linear
response. This method also allows to assess corrections from the
\emph{energy dependence of the mixing angles in the medium} upon the
wave-packet dynamics. }

\item{iii:) The space-time evolution of the initially prepared
wave packet, including dispersive effects allows an assessment of
the competition between the  progressive loss of coherence in the
wave-packet dynamics    by the separation of mass eigenstates,
collisional decoherence and cosmological expansion. While our study
only includes the self-energy from charged and neutral currents up
to one-loop, the final result allows us to include results available
in the literature\cite{barbieri,enqvist,notzold,dolgov} to
understand the effects of collisional decoherence and cosmological
expansion when there is a separation of time scales.  }

\item{iv): We focus our study within the context of early Universe
cosmology, in particular in the temperature regime just prior to BBN
where there is a possibility for resonant
transitions\cite{barbieri,enqvist,notzold,dolgov,hoboya}. Of
particular interest are the medium modifications of the dispersion
relations, wave-packet dispersion, oscillation and coherence time
scales in this temperature and energy regime. }

\item{v): A bonus of this field theoretical formulation is that it
also allows to obtain the space-time evolution of right-handed
neutrino wave packets. Although the amplitude for such wave packets
is suppressed by $M/k$ with $M,k$ the typical  mass and energy scale
of neutrinos in the medium,  and may not be relevant for neutrino
processes in the early Universe, the method systematically yields
this information. }

\end{itemize}

Since we   study  the propagation of neutrino wave-packets in a
medium, aspects   associated with the source-detector measurement
processes are not well-defined   or relevant in this case.
Consequently, in contrast to most studies in the literature,
\emph{we do not integrate in time} as is the case for a description
of experiments in the vacuum \cite{CG1998}. Therefore, our study of
propagation is both in \emph{space and time}.

\subsection{  Main results:}

Our main results are the following:

\begin{itemize}
\item{ A systematic field theoretical formulation of the
space-time dynamics of wave packets of massive and mixed neutrinos
in terms of the effective Dirac equation including self-energy
corrections. The space-time evolution is approached as an initial
value problem via linear response with the full density matrix. }

\item{Wave packet evolution features   characteristic time
scales associated with transverse and longitudinal dispersion. The
ratio of these   scales is given by the enormous Lorentz dilation
factor in the case of relativistic neutrinos. Neither of these
scales receives substantial medium corrections. The shortest scale,
associated with the transverse dispersion dominates the suppression
of both the survival and transition probabilities. There is an
interesting phenomenon of ``coherence freezing'' which results from
the competition between longitudinal dispersion and the separation
of the mass eigenstates. We find that there are medium as well as
wave-packet modifications to the oscillation formula both for the
oscillation frequency as well as the survival and transition
probabilities. }

\item{There is a resonance for the mixing angle in the medium just
prior to BBN\cite{barbieri,notzold,enqvist,dolgov,hoboya} at an
energy and temperature $k\sim T \sim 3.6\, \textrm{MeV}$ for large
mixing angle or $k\sim T \sim 7\, \textrm{MeV}$ for small mixing
angle. Both the coherence  and the oscillation time scales are
enhanced by a factor $1/\sin2\theta$ near the resonance, where
$\theta$ is the \emph{vacuum} mixing angle. This suggests a
substantial increase both in the coherence and the oscillation
scales for $1-3$ mixing, but not an appreciable modification for
$1-2$ mixing. }

\item{Assuming that the momentum spread of the initial wavepacket
is determined by the large angle scattering mean free path of
charged leptons\cite{book1}, we find that near the resonance the
loss of coherence via charged and neutral current elastic scattering
is faster than the loss of coherence by the separation of the mass
eigenstates and occurs on a time scale much shorter than the Hubble
time. However, the decoherence time scale is many orders of
magnitude \emph{larger} than the time scale for transverse
dispersion which ultimately determines the suppression of the
survival and transition probabilities. }
\end{itemize}

\subsection{Outline:}

This article is organized as follows. In section \ref{sec:Dirac}, we
obtain the effective Dirac equation of neutrino in a thermal medium
implementing the methods of non-equilibrium field theory and linear
response\cite{hoboya}. In this section we obtain the in-medium
dispersion relations and mixing angles. In section
\ref{sec:wave-packet}, we develop the general formulation to study
the space-time propagation of neutrino wave-packet.  In this section
we discuss the different time scales associated with dispersion,
oscillations and coherence.  In section \ref{sec:timescales} we
compare the different time scales with the Hubble and collisional
relaxation time scales and discuss the impact of the different
scales upon the space-time evolution of the neutrino wave-packets,
coherence and oscillations. We present our   conclusions in section
\ref{sec:DC}.

\section{Effective Dirac equation of Neutrinos in a medium and linear response}
\label{sec:Dirac}

The study of the evolution of neutrino wave packets in the vacuum
typically involves a description of the experimental production and
detection of these wave packets.  We study the space-time evolution
of wave packets in a medium as an \emph{initial value problem}. This
is achieved in \emph{linear response} by coupling an external source
term which induces an expectation value of  the neutrino field in
the density matrix, after this source is switched off the
expectation value evolves in time. The propagation of this initial
state is described by the effective Dirac equation in the medium,
which includes the self-energy corrections. This is the familiar
linear response approach to studying the evolution out of
equilibrium in condensed matter systems. The correct framework to
implement this program is the real-time formulation of field theory
in terms of the closed-time-path integral
\cite{ctp,disip,nosfermions}.

We restrict our study to the case of two Dirac flavor neutrinos,
while the formulation is general and can treat $1-2$ or $1-3$ mixing
on equal footing, for convenience we will  refer to  electron and
muon neutrino mixing. Neutrino mixing and oscillations is
implemented by adding to the Standard Model a Dirac mass matrix
$M_{ab}$ which is off-diagonal in the flavor basis. For our
discussion, the relevant part of the Lagrangian is given by

\be\label{LSM} \mathcal{L} = \mathcal{L}^0_{\nu} +
\mathcal{L}^0_W+\mathcal{L}^0_Z + \mathcal{L}_{CC}+\mathcal{L}_{NC}+
\bar{\eta}_a \nu_a + \bar{\nu}_a \eta_a,\ee

\noindent where $\mathcal{L}^0_{\nu}$ is the free field neutrino
Lagrangian density

\be\label{FFnuL} \mathcal{L}^0_{\nu} = \overline{\nu}_a \left(i
{\not\!{\partial}}\,\delta_{ab}-M_{ab}\right) \nu_b \ee

\noindent where $a, b$ are flavor indices. $\mathcal{L}^0_{W,Z}$ are
the free field vector boson Lagrangian densities in the unitary
gauge, namely

\bea \mathcal{L}^0_W &=& -\frac12
\left(\partial_{\mu}W^+_{\nu}-\partial_{\nu}W^+_{\mu}
\right)\left(\partial^{\mu}W^{-\,\nu}-\partial^{\nu}W^{-\,\mu}
\right)+ M^2_W \; W^+_{\mu} \; W^{-\,\mu} \; , \label{LW}\\
\mathcal{L}^0_Z & = & -\frac{1}{4}
\left(\partial_{\mu}Z_{\nu}-\partial_{\nu}Z_{\mu}
\right)\left(\partial^{\mu}Z^{\nu}-\partial^{\nu}Z^{\mu} \right)+
\frac12M^2_Z \; Z_{\mu} \; Z^{\mu} \; , \label{LZ} \eea

\noindent and the charged and neutral current interaction Lagrangian
densities are given by

\be \mathcal{L}_{CC} = \frac{g}{\sqrt2} \left[ \overline{\nu}_a \;
\gamma^\mu \;  L  \;  l_a \;  W^+_{\mu} + \overline{l}_a \gamma^\mu
\;  L \;  \nu_a \;  W^-_{\mu} \right] \; ,\label{LCC} \ee \be
\mathcal{L}_{NC} = \frac{g}{2 \cos \theta_w} \left[ \overline{\nu}_a
\; \gamma^\mu \;  L \;  \nu_a \;  Z_{\mu} + \overline{f}_a \;
\gamma^\mu \;  (g^V_a-g^A_a\,\gamma^5) \;  f_a \; Z_{\mu}
\right]\label{LNC} \; , \ee

\noindent where $\theta_w$ is the Weinberg angle, $L=(1-\gamma^5)/2$
is the left-handed chiral projection operator, and $g^{V,A}$ are the
vector and axial vector couplings for quarks and leptons. The label
$l$ stands for leptons and $f$ generically for the fermion species
with neutral current interactions. The external sources
$\eta_a,\bar{\eta}_a$ which couple to the neutrino fields depend
explicitly on space and time and induce an expectation value whose
time evolution is studied in linear response.

For two Dirac flavor neutrinos, the mass matrix $M_{ab}$ is
parametrized by

\be \label{massmatrix} \mathds{M}=\left(%
\begin{array}{cc}
  m_{ee} & m_{e\mu} \\
  m_{e\mu} & m_{\mu\mu} \\
\end{array}%
\right) . \ee

For the vacuum case, the elements $ m_{ee}, m_{\mu\mu} $ and $
m_{e\mu} $ are related to the vacuum mixing angle $\theta$ and
masses of the propagating mass eigenstates $ M_1 $ and $ M_2$ as
follows

\bea m_{ee}=\cos^2\theta \; M_1+\sin^2\theta \;
M_2~;~~m_{\mu\mu}=\sin^2\theta \; M_1+\cos^2\theta \;
M_2~;~~m_{e\mu}=-(M_1-M_2)  \sin\theta  \cos\theta \; . \eea

For later convenience and to establish contact with observable
parameters  we introduce the following quantities

\be \overline{M}= \frac{M_1+M_2}{2} ~~;~~ \delta M^2 = M^2_1-M^2_2
\label{mbardm}, \ee

The masses $M_{1,2}$ can be conveniently written in terms of these
quantities as

\bea \label{Ma} M_a=\OM\left[ 1+(-1)^{a-1}\frac{\delta
M^2}{4\OM^2}\right]~;~~ a=1,2 \,. \eea

The current value for $\OM$ obtained by WMAP \cite{WMAP} and the
oscillation parameters from the combined fitting of the solar and
KamLAND data are \cite{kamland} respectively:

\be \label{data}\overline{M} \approx 0.25 \,(eV)~~;~~|\delta
M^2_{12}| \approx 7.9\times 10^{-5}\,(eV)^2~;~~\tan^2\theta_{12}
\approx 0.40 \; \,. \ee

For atmospheric neutrinos, analysis from SuperKamiokande, CHOOZ and
atmospheric neutrino data yield,

\be |\delta M^2_{13}| \approx (1.3-3.0)\times
10^{-3}\,(eV)^2~;~~\sin^2\theta_{13} < 0.067  ~ (3\sigma)\; \,. \ee

\no This implies that $|\delta M^2|/\,\overline{M}^{\,2} \ll 1 $, an
almost degenerate hierarchy of neutrino masses.

\subsection{Linear response:}

The medium is described by an \emph{ensemble} of states, and the
description is in terms of a density matrix. Therefore the question
of space-time evolution is more subtle, while in the vacuum one can
consider preparing an initial \emph{single particle} state and
evolving it in time, such a consideration is not available in a
medium, and the question of time evolution \emph{must} be formulated
differently, namely in terms of expectation values of the relevant
operators in the density matrix.

In equilibrium the neutrino field \emph{cannot} have an expectation
value in the density matrix. The usual method in many body theory to
study the non-equilibrium evolution of single quasiparticle states
is the method of linear response: an external source is coupled to
the field which develops an expectation value in the density matrix
induced by the source.   The expectation value of this field obeys
the equation of motion with medium corrections. Upon switching-off
the external source, the expectation value evolves in time as a
solution of the effective equations of motion in the medium. For a
detailed description of this method in non-equilibrium quantum field
theory see refs.\cite{disip,nosfermions,hoboya}. The external
sources $\eta_a$ in the Lagrangian density (\ref{LSM}) induce an
expectation value of the neutrino field \be   \psi_a = \langle \nu_a
\rangle = \mathrm{Tr}\,\hat{\rho}\,\nu_a \label{shift}\ee where
$\hat{\rho}$ is the full density matrix of the medium. In linear
response this expectation value is linear in the external source and
obeys the effective Dirac equation of motion in the
medium\cite{nosfermions}. It is most conveniently written in terms
of the spatial Fourier transforms of the fields, sources and
self-energies $\psi_a(\vk,t),\eta_a(\vk,t),\Sigma(\vk,t-t')$
respectively\footnote{We have kept the same functions to avoid
cluttering the notation, but the label $\vk$ makes it clear that
these are the spatial Fourier transforms.}. The one loop
self-energies with neutral and charged current contributions had
been obtained in refs.\cite{notzold,enqvist,hoboya}, and the
effective Dirac equation in the medium up to one loop has been
obtained in the real time formulation in ref.\cite{hoboya}. It is
given by

\be \left[\left( i\gamma^0 \frac{\partial}{\partial
t}-\vec{\gamma}\cdot \vec{k} \right)
\,\delta_{ab}-M_{ab}+\Sigma^{tad}_{ab}L\right]\,\psi_b (\vk,t) +
 \int_{-\infty}^t dt' \Sigma_{ab}(\vk,t-t')\,L \,\psi_b(\vk,t') =
- \eta_a(\vk,t), \label{eqnofmotft}\ee where $L$ is the projector on
left handed states, $\Sigma^{tad}_{ab}L$ is the (local) tadpole
contribution from the neutral currents and $\Sigma _{ab}(\vk,t-t')$
is the spatial Fourier transform of the (retarded)
 self energy which includes both neutral and charged current
interactions, and whose spectral representation is given by

\be \Sigma (\vk,t-t') = i \,\int_{-\infty}^\infty \frac{dk_0}{\pi}\,
\mathrm{Im}\Sigma(\vk,k_0)\, e^{-ik_0(t-t')} ~~;~~
\Sigma(\vk,k_0)=\Sigma_W(\vk,k_0)+\Sigma_Z(\vk,k_0)\label{selfa}\ee
where we have separated the charged and neutral current
contributions respectively.

The external source term $\eta$ allows to ``prepare'' a determined
initial state, leading to the time evolution of $\psi$ as an initial
value problem. This approach is implemented as follows. Consider
switching on the source adiabatically from $t=-\infty$ up to $t=0$
and switching it off at $t=0$, \be \eta_a(\vk,t) =
\eta_a(\vk,0)\,e^{\epsilon t} \theta(-t)~~;~~\epsilon \rightarrow
0^+ \ee It is straightforward to confirm that for $t < 0$ the
solution of the Dirac eqn. (\ref{eqnofmotft}) is given by

\be  \psi_a(\vk,t<0) =  \psi_a(\vk, 0)\,e^{\epsilon\,t}
\label{solless} \ee

Inserting this ansatz into the equation (\ref{eqnofmotft}) yields  a
linear relation which determines the initial value $\psi_a(\vk, 0)$
from  $\eta_a(\vk,0)$, or equivalently, for a given initial value
$\psi(\vk,0)$ allows to determine the adiabatic source that yields
this initial value problem. The evolution for $t>0$ is determined by
the following effective (retarded) Dirac equation,

\be \left[\left( i\gamma^0 \frac{\partial}{\partial
t}-\vec{\gamma}\cdot \vec{k} \right)
\,\delta_{ab}-M_{ab}+\Sigma^{tad}_{ab}L\right]\,\psi_b (\vk,t) +
 \int_{0}^t dt' \Sigma_{ab}(\vk,t-t')\,L \,\psi_b(\vk,t') = -\psi_a(\vk,
 0)\,\int_{-\infty}^\infty \frac{dk_0}{\pi}\,
\frac{\mathrm{Im}\Sigma(\vk,k_0)}{k_0} \, e^{-ik_0 t } ,
\label{eqntgreat0}\ee

This equation can be solved by Laplace transform. Introduce the
Laplace transforms

\be \label{laplatrafo} \widetilde{\psi}_b(\vk,s) = \int_0^\infty
e^{-st} \psi_b(\vk,t )~~;~~ \widetilde{\Sigma}(\vk,s) =
\int_0^\infty e^{-st} \Sigma(\vk,t ) = \int_{-\infty}^\infty
\frac{dk_0}{\pi}\, \frac{\mathrm{Im}\Sigma(\vk,k_0)}{k_0-is} \ee

\noindent where we have used eqn. (\ref{selfa}) to obtain the
Laplace transform of the self-energy. At this stage it is convenient
to establish contact with the more familiar description of the Dirac
equation in the frequency representation (Fourier transform in time)
by introducing the time Fourier transform of the retarded self
energy,

\be \Sigma(\vk,\omega) = \int \frac{dk_0}{\pi}
\frac{\textrm{Im}\Sigma(\vk,k_0)}{k_0-\omega -i\epsilon}\,,
\label{SigFT}\ee related via analyticity to the Laplace transform,
namely \be \widetilde{\Sigma}(\vk,s) =
\Sigma(\vk,\omega=is-i\epsilon) \label{analit} \ee

In terms of Laplace transforms the equation of motion becomes the
following algebraic equation

\be D_{ab}(\vk,s)\, \widetilde{\psi}_b(\vk,s) =
i\left(\gamma^0\,\delta_{ab}+\frac{1}{is}\left[\widetilde{\Sigma}_{ab}(\vk,s)-\widetilde{\Sigma}_{ab}(\vk,0)\right]\,L\right)\psi_b(\vk,0)
\label{lapladirac} \ee where $D(\vk,s)\equiv
D(\vk,\omega=is-i\epsilon)$ is the analytic continuation of the
Dirac operator in frequency and momentum space   \be
D_{ab}(\vk,\omega) = \left[ \left(  \gamma^0
\omega-\vec{\gamma}\cdot \vec{k} \right)
\,\delta_{ab}-M_{ab}+\Sigma^{tad}_{ab}\,L   +
 {\Sigma}_{ab}(\vk,\omega)\,L \right] \label{Dirlapla}
\ee

The full space-time evolution of an initial state is determined by

\be \psi_a(\vx,t) = \int  {d^3k}\,e^{i\vk\cdot\vx}\, \int_{\Gamma}
\frac{ds}{2\pi i}\,D^{-1}_{ab}(\vk,s)\,(i\gamma^0) \psi_b(\vk,0)\,
e^{st} \; ,\label{inverlapla} \ee \noindent where $\Gamma$ is the
Bromwich contour in the complex $s$ plane running parallel to the
imaginary axis to the right of all the singularities of the function
$\widetilde{\psi}(\vk,s)$ and closing on a large semicircle to the
left. We have simplified the expression for the eqn.
(\ref{inverlapla}) by discarding a perturbatively small correction
to the amplitude of $\mathcal{O}(G_F)$, given by the self-energy
corrections on the right hand side of eqn. (\ref{lapladirac}).
Therefore the space-time evolution is completely determined by the
singularities of the function $\widetilde{\psi}(\vk,s)$ in the
complex s-plane. Up to one loop order and for temperatures well
below the mass of the vector bosons, the only singularities are
simple poles along the imaginary axis, corresponding to the
dispersion relations of the propagating modes in the medium. In this
temperature range absorptive processes emerge at the two loop level,
consequently these are of $\mathcal{O}(G^2_F)$ and are neglected in
the one loop  analysis presented here. The integral along the
Bromwich contour in the complex s-plane can now be written

\be \int_{\Gamma} \frac{ds}{2\pi i}\,D^{-1}_{ab}(\vk,s)\,(i\gamma^0)
\psi_b(\vk,0)\, e^{st} = \int_{-\infty}^{\infty} \frac{d\omega}{2\pi
 }\,D^{-1}_{ab}(\vk,\omega)\,(i\gamma^0) \psi_b(\vk,0)\,
 e^{-i\om\,t}\label{integ} \ee where the frequency integral is performed along a
 line parallel to but slightly below the real $\omega$ axis
 closing counterclockwise in the upper half plane.


The one loop contributions to the self-energy for $\omega,k,T \ll
M_W$ were obtained in reference\cite{notzold,enqvist,hoboya} and
found to be of the form\cite{hoboya}\be
\Sigma^{tad}_{ab}+\Sigma_{ab}(\vk,\omega) = \gamma^0 \mathds{A}(\om
)-\vec{\ga}\cdot\uvk \,\mathds{B}( k)\ee where $\mathds{A}(\om )$
and $\mathds{B}(k)$ are $2\times2$ diagonal matrices in the neutrino
flavor basis.\footnote{The equivalence with the notation of
ref.\cite{notzold} is (see eqn. (2) in ref.\cite{notzold}): $a_{NR}
= \mathds{B}( k)/k~~;~~ b_{NR}= \mathds{A}(\om )-\omega
\,\mathds{B}( k)/k$. }. To lowest order in $k/M_{W}\,;\omega/M_W$
these matrices are found to be\cite{notzold,enqvist,dolgov,hoboya}

\bea \mathds{A}(\om )= \left(
                   \begin{array}{cc}
                    A_e(\om ) & 0 \\
                    0& A_\mu(\om ) \\
                   \end{array}
                 \right)~;~~
\mathds{B}(k) = \left(
                   \begin{array}{cc}
                    B_e( k) & 0 \\
                    0& B_\mu( k) \\
                   \end{array}
                 \right).
\eea

Imposing charge neutrality, the results of reference\cite{hoboya}
(see also\cite{notzold,enqvist,barbieri,dolgov}) are summarized in
the following regimes: high temperature and density, relevant during
the early Universe or low temperature and high density, relevant for
cold dense matter in supernovae, neutron stars or the sun
\begin{itemize}
\item{  $\mathbf{m_e \ll T \ll m_\mu}$

In this regime we consider the following degrees of freedom $\nu,e$
and $  p,n$ in nuclear statistical equilibrium. The matrix elements
are given by

\bea  A_e(\om )&=& \frac{G_F\;n_\ga}{\sqrt2}\Bigg[-\mathcal{L}_e
+\frac{7 \; \pi^4}{60 \; \zeta(3)} \; \frac{\om \;
T}{M_W^2}\left(2+\cos^2\theta_w \right)\Bigg] \; , \label{Aelo} \\
A_\mu(\om )&=& \frac{G_F \; n_\ga}{\sqrt2}\Bigg[-\mathcal{L}_\mu
+\frac{7 \; \pi^4}{60 \; \zeta(3)} \; \frac{\om \;
T}{M_W^2}\cos^2\theta_w \Bigg] \; ,
\label{Aulo}\\
B_e( k)&=& -\frac{G_F\;n_\ga}{\sqrt2} \frac{7 \; \pi^4}{180 \;
\zeta(3)} \; \frac{k \;
T}{M_W^2}\left(2+ \cos^2\theta_w \right) \; ,\label{Belo} \\
B_\mu( k)&=& -\frac{G_F\;n_\ga}{\sqrt2}\frac{7 \; \pi^4}{180 \;
\zeta(3)} \; \frac{k \; T}{M_W^2}\cos^2\theta_w \;  , \label{Bulo}
\eea where in terms of the   asymmetries $L_f$ with $f=\nu,e, p,n$

\bea -\mathcal{L}_e  =  -\frac12 \; L_{\nu_e}+L_{\nu_{\mu}}-3 \;
L_e-L_n ~;~~ -\mathcal{L}_\mu  = -\frac12L_{\nu_{\mu}}+L_{\nu_e}-L_n
\label{els} \eea }

\item{  $\mathbf{m_e,m_\mu \ll T \ll M_W}$

In this regime the relevant degrees of freedom are the lightest
deconfined quarks $u,d$ and leptons

\bea A_e(\om )&=& \frac{G_F\,n_\ga}{\sqrt{2}}\Bigg[\
\widetilde{-\mathcal{L}_e} +\frac{7\pi^4}{60\zeta(3)}\frac{\om
T}{M_{W}^2}\left(2+ \cos^2\theta_w \right) \Bigg]\label{Aehi} \\
 A_\mu(\om )&=&
\frac{G_F\,n_\ga}{\sqrt{2}}\Bigg[-\widetilde{\mathcal{L}_\mu} +
\frac{7\pi^4}{60\zeta(3)}\frac{\om
T}{M_{W}^2}\left(2+ \cos^2\theta_w \right)\Bigg]\label{Auhi} \\
B_e( k)&=& -\frac{G_F\,n_\ga}{\sqrt{2}}
\frac{7\pi^4}{180\zeta(3)}\frac{k T}{M_{W}^2}\left(2+ \cos^2\theta_w
-
\frac{60}{7\pi^2}\left(\frac{m_e}{T}\right)^2\right)\label{Behi}\\
B_\mu( k)&=& -\frac{G_F\,n_\ga}{\sqrt{2}}
\frac{7\pi^4}{180\zeta(3)}\frac{k T}{M_{W}^2}\left(2+ \cos^2\theta_w
- \frac{60}{7\pi^2}\left(\frac{m_\mu}{T}\right)^2\right).
\label{Buhi} \eea

\noindent where

\bea -\widetilde{\mathcal{L}_e} & = &
-\frac{1}{2}L_{\nu_e}+L_{\nu_{\mu}}-3L_e +
( 1-4\sin^{2}\theta_w)(2L_e-L_\mu)-(1-8\sin^{2}\theta_w)L_u-2L_d  \\
-\widetilde{\mathcal{L}_\mu}  & = &
-\frac{1}{2}L_{\nu_{\mu}}+L_{\nu_e}-3L_\mu + (
1-4\sin^{2}\theta_w)(2L_e-L_\mu)-(1-8\sin^{2}\theta_w)L_u-2L_d
\label{widels}\eea  }

\item{ Cold dense matter with $e,\nu,\mu,p,n$ :

\bea A_e &=& -\frac{G_F\,\mathcal{N}_e}{\sqrt{2}} ~~;~~ A_\mu  =
-\frac{G_F\,\mathcal{N}_\mu}{\sqrt{2}} ~~;~~ B_{e,\mu}   =   0 \\
-\mathcal{N}_e   & = & -\frac12 \;
\mathcal{N}_{\nu_e}+\mathcal{N}_{\nu_{\mu}}-3 \;
\mathcal{N}_e-\mathcal{N}_n \\ -\mathcal{N}_\mu  & = & -\frac12
\mathcal{N}_{\nu_{\mu}}+\mathcal{N}_{\nu_e}-\mathcal{N}_n \eea }

\item{ Cold dense matter with quarks and leptons :

\bea A_e &=& -\frac{G_F }{\sqrt{2}}\,\widetilde{\mathcal{N}}_e ~~;~~
A_\mu  = -\frac{G_F }{\sqrt{2}}\,\widetilde{\mathcal{N}}_\mu~~;~~
B_{e,\mu} =
0 \\
-\widetilde{\mathcal{N}_e} & = &
-\frac{1}{2}\mathcal{N}_{\nu_e}+\mathcal{N}_{\nu_{\mu}}-3\mathcal{N}_e
+ (
1-4\sin^{2}\theta_w)(2\mathcal{N}_e-\mathcal{N}_\mu)-(1-8\sin^{2}\theta_w)\mathcal{N}_u-2\mathcal{N}_d
\\
-\widetilde{\mathcal{N}_\mu}  & = &
-\frac{1}{2}\mathcal{N}_{\nu_{\mu}}+\mathcal{N}_{\nu_e}-3\mathcal{N}_\mu
+ (
1-4\sin^{2}\theta_w)(2\mathcal{N}_e-\mathcal{N}_\mu)-(1-8\sin^{2}\theta_w)\mathcal{N}_u-2\mathcal{N}_d\eea
}

\end{itemize}

\no where $n_\ga=2\,\zeta(3)\, T^3/\pi^2$ is the photon density, and
\be L_f= \frac{\mathcal{N}_f}{n_\gamma}~~;~~\mathcal{N}_f= 2 \int
\frac{d^3p}{(2\pi)^3} \left[N_f(p)-N_{\bar f}(p)\right]\,,  \ee
$\mathcal{N}_f$ are the particle-antiparticle asymmetry
\emph{densities}.

\subsection{Dispersion Relations and Mixing Angles in the
Medium}\label{disprel}

The  simple poles of the integrand in (\ref{integ}) are the
solutions of the homogeneous Dirac equation \be \label{DiracEq}
\left[\gamma^0\,\omega\mathds{1}- \vec{\gamma}\cdot \uvk\, k
\mathds{1}-\mathds{M}+ \left(\gamma^0 \mathds{A}(\om
)-\vec{\ga}\cdot\uvk \,\mathds{B}( k) \right)L \right]\,\psi
(\omega,k)= 0 \; , \ee \no where $\mathds{1}$ is the $2\times2$
identity matrix in the flavor basis in which the field $\psi(\om,k)$
is given by \be \psi(\om,k)=\left(
                  \begin{array}{c}
                    \nu_e(\om,k) \\
                    \nu_{\mu}(\om,k) \\
                  \end{array}
                \right) \; ,
\ee \no with $\nu_e(\om,k)$ and $\nu_{\mu}(\om,k)$ each being a
4-component Dirac spinor.

It turns out to be most convenient to work in the chiral basis in
which the left-handed and right-handed components of the Dirac
doublets are written as \cite{hoboya}

\be \psi_L = \sum_{h=\pm1} \left(
                        \begin{array}{ c}
                            0 \\
                          v^{(h)}\otimes \varphi^{(h)} \\
                        \end{array}
                      \right) ~~;~~ \psi_R = \sum_{h=\pm1} \left(
                        \begin{array}{ c}
                            v^{(h)}\otimes \xi^{(h)} \\
                          0 \\
                        \end{array}
                      \right) \label{heli}\ee

\no where the two component Weyl spinors $v^{(h)}$ are the
eigenstates of the helicity operator $\vec{\sigma}\cdot \uvk$ with
eigenvalues $h=\pm 1$.

To the leading order in $G_F$, the left-handed flavor doublet

\be \varphi^{(h)} (\omega,k) = \left(
         \begin{array}{c}
           \nu^{(h)}_{e } (\omega,k)\\
           \nu^{(h)}_{\mu  }(\omega,k) \\
         \end{array}
       \right) \label{flavdou} \; , \ee

\no obeys the following effective Dirac equation \cite{hoboya}

\be \label{eignEqL} \left[(\om^2-k^2)\mathds{1}+(\om-h
k)(\mathds{A}+h \mathds{B})- \mathds{M}^2 \right]\varphi^{(h)}
(\omega,k)=0 \; , \ee

\no while the right-handed doublet is determined by the relation
\cite{hoboya}

\be \label{xieq} \xi^{(h)}(\omega,k) = -  \; \frac{(\om +h \; k)
}{\omega^2-k^2} \; \mathds{M}\, \varphi^{(h)}(\omega,k) \; . \ee

The propagating modes in the medium are found by diagonalization of
the above effective Dirac equation. This can be done by performing a
unitary transformation  $\varphi^{(h)}(\omega,k) = U^{(h)}_m \;
\chi^{(h)}(\omega,k)$ where

\bea \label{mediumrotmat} U^{(h)}_m = \left(
          \begin{array}{cc}
            \cos\theta_m^{(h)} & \sin \theta_m^{(h)} \\
            -\sin \theta_m^{(h)} & \cos \theta_m^{(h)} \\
          \end{array}
        \right) \; ~;~~\label{massei} \chi^{(h)}(\om,k) = \left(
                  \begin{array}{c}
                    \nu^{(h)}_1(\om,k) \\
                    \nu^{(h)}_2(\om,k) \\
                  \end{array}
                \right) \; ,
\eea

\no and a similar transformation for the right handed doublet
$\xi^{(h)}(\omega,k)$,  with the medium mixing angle
$\theta^{(h)}_m$ depending on  $h,k$ and $\omega$. Upon
diagonalization, the eigenvalue equation is given by \cite{hoboya}

\be  \Bigg\{ \omega^2-k^2 +\frac12 \;  S_h(\omega,k)-\frac12 \;
(M^2_1+M^2_2)-\frac12 \; \delta\,M^2 \, \Om_h(\om,k)\left(
          \begin{array}{cc}
           1 & 0 \\
            0 & -1 \\
          \end{array}
        \right)\Bigg\}\chi^{(h)}(\omega,k)
          =0 \label{psimass}  \; ,
\ee

\no where $ S_h(\omega,k), \,\Delta_h(\om,k)$ and $\Om_h(\om,k)$ are
respectively given by

\bea S_h(\omega,k) & = & (\om-h k)\left[A_e(\omega )+A_\mu(\omega )
+h  \; B_e( k)+h \; B_\mu( k)\right]
\label{Sofome} \; , \\
\Delta_h(\omega,k) &  = & (\om-h k)\left[A_e(\omega )-A_\mu(\omega
)+h \; B_e( k)-h \; B_\mu( k) \right]\label{Deltah} \;, \\
\Om_h(\om,k) &=& \left[\left(\cos 2\theta-\frac{\Delta_h
(\om,k)}{\delta M^2}\right)^2+\sin^22\theta
\right]^{\frac12}\label{Omegah}\;. \eea

\no This requires the matrix elements in $U^{(h)}_m $ to be of the
following form

\bea \label{SinCos}\sin 2\theta^{(h)}_m(\om ,k)
=\frac{\sin2\theta}{\Om_h(\om,k)}~;~~\cos 2\theta^{(h)}_m(\om,k)
  =\frac{\cos 2\theta-\frac{\Delta_h(\om,k)}{\delta
M^2}}{\Om_h(\om,k)}. \eea

A resonance is available whenever

\be \left(\cos 2\theta-\frac{\Delta_h (\om,k)}{\delta M^2}\right)
\approx 0 \label{resonance}\ee in which case $\sin
2\theta^{(h)}_m(\om ,k)\approx 1~~;~~ \cos 2\theta^{(h)}_m(\om
,k)\approx 0$.

The solutions of eqn. (\ref{psimass}) yield the dispersion relations
$\omega(k)$  of the ``exact'' quasiparticle states in the medium and
correspond to the ``exact poles'' of the Dirac propagator. The
dispersion relations   $\om_a^{(h)}(k,\la)$ for the propagating
modes in the medium are found in perturbation theory consistently up
to $\mathcal{O}(G_F)$ by writing \cite{hoboya}

\be  \omega^{(h)}_a(k,\la) = \la
\left[E_a(k)+\delta\omega^{(h)}_a(k,\la)\right] \quad , \quad
a=1,2~~;~~ \lambda = \pm \label{omegaa} \ee

\no where $ E_{a}(k) = \sqrt{k^2+M^2_{a}}$\,, and
$\delta\omega^{(h)}_a(k,\la)$ are found to be

\bea \delta\omega^{(h)}_a(k,\la)= -\frac{1}{4E_a(k)}\Bigg\{S_h(\la
E_a(k),k)+(-1)^a\,\delta M^2 \,\left(\,\Om_h(\la
E_a(k),k)-1\,\right) \Bigg\}. \eea

For relativistic neutrinos with $k \gg M_a$ the dispersion relations
$\omega_a(k)~~;~~a=1,2$ for the different cases are given to leading
order in $G_F$     by

\begin{itemize}
\item{{\bf Positive energy, negative helicity neutrinos}, $\lambda=+1,
h=-1$:

\bea \label{posenegh} \omega_a(k) = k+
\frac{M^2_a}{2k}-\frac{1}{4k}\Bigg[S_{-}(k,k)+(-1)^a\delta M^2
\Big(\Omega_{-}(k,k)-1 \Big) \Bigg]  \,. \eea

 }

\item{{\bf Positive energy, positive helicity neutrinos}, $\lambda=+1,
h=+1$:

\bea \omega_a(k) = k+
\frac{M^2_a}{2k}-\frac{1}{4k}\Bigg[S_{+}(k,k)+(-1)^a\delta M^2
\Big(\Omega_{+}(k,k)-1 \Big) \Bigg] ~~;~~\omega - h\,k \approx
\frac{\overline{M}^2}{2k}   \eea where we have neglected corrections
of order $\delta M^2/\overline{M}^2$.

 }

 \item{{\bf Negative energy, negative helicity neutrinos}, $\lambda=-1,
h=-1$:

\bea  \omega_a(k) = -k-
\frac{M^2_a}{2k}+\frac{1}{4k}\Bigg[S_{-}(-k,k)+(-1)^a\delta M^2
\Big(\Omega_{-}(-k,k)-1 \Big) \Bigg]  ~~;~~ \omega -h\,k \approx
\frac{\overline{M}^2}{2k}   \eea where we have again neglected
corrections of order $\delta M^2/\overline{M}^2$.

 }

  \item{{\bf Negative energy, positive helicity neutrinos}, $\lambda=-1,
h=+1$:

\bea  \omega_a(k) = -k-
\frac{M^2_a}{2k}+\frac{1}{4k}\Bigg[S_{+}(-k,k)+(-1)^a\delta M^2
\Big(\Omega_{+}(-k,k)-1 \Big) \Bigg] \,. \eea

 }

\end{itemize}

In the above expressions  the $\Omega_\pm$ are given by eqn.
(\ref{Omegah}).

The vacuum and medium oscillation time scales are respectively
defined as

\bea T_{vac} &=& \frac{2\pi}{ {E}_1- {E}_2}~;~~ T_{med} =
\frac{2\pi}{  \omega^{(h)}_1(k,\lambda)-\omega^{(h)}_2(k,\lambda) },
\eea

\no In  the relativistic case when $k\gg M_a$, we find  \be
 T_{vac}\app \frac{4\pi k}{\delta M^2}~~;~~ T_{med} \approx \frac{4\pi k}{\delta
M^2\,\Om_h(\lambda\,k,k)}
 \ee leading to the relation

 \be \label{reltimes} \frac{T_{med}}{T_{vac}} = \frac{\sin 2\theta^{(h)}_m(\om
 ,k) }{\sin 2\theta } \ee

\section{Space-time propagation of a neutrino wave-packet.}
\label{sec:wave-packet}

We now have all the ingredients necessary to study the space-time
evolution of a initial wave packet by carrying out the integrals in
eqn. (\ref{inverlapla}). For this purpose it is convenient to write
$\psi(\vk,0) = \psi_R(\vk,0)+\psi_L(\vk,0)$ and expand the right and
left-handed components in the helicity basis as in eqn.
(\ref{heli}), namely

\be \psi_L(\vk,0) = \sum_{h=\pm1} \left(
                        \begin{array}{ c}
                            0 \\
                          v^{(h)}\otimes \varphi^{(h)}(\vk,0) \\
                        \end{array}
                      \right) ~~;~~ \psi_R(\vk,0) = \sum_{h=\pm1} \left(
                        \begin{array}{ c}
                            v^{(h)}\otimes \xi^{(h)}(\vk,0) \\
                          0 \\
                        \end{array}
                      \right) \label{heli2}\ee where

 \be \varphi^{(h)}(\vk,0) = \left(
                        \begin{array}{ c}
                            \nu^{(h)}_{eL}(\vk,0) \\
                          \nu^{(h)}_{\mu L}(\vk,0) \\
                        \end{array} \right)~~;~~ \xi^{(h)}(\vk,0) = \left(
                        \begin{array}{ c}
                            \nu^{(h)}_{e R}(\vk,0) \\
                          \nu^{(h)}_{\mu R}(\vk,0) \\
                        \end{array} \right) \ee

The general initial value problem requires to furnish initial
conditions for the four components above. However, an inhomogeneous
neutrino state is produced by a weak interaction vertex, which
produces left handed neutrinos, suggesting to set
$\nu^{(h)}_{eR}(\vec{k},0)=0;\nu^{(h)}_{\mu R}(\vec{k},0)=0$.
Without loss of generality let us consider an initial state
describing  an inhomogeneous wave-packet of \emph{electron
neutrinos} of arbitrary helicity, thus $\nu^{(h)}_{eL}(\vk,0)\neq
0;\nu^{(h)}_{\mu L}(\vk,0)=0$.

In the cases of interest neutrinos are relativistic with typical
momenta $k\gg \overline{M}$. Following the real time analysis
described in detail in ref.\cite{hoboya} in the relativistic case we
find

\bea \label{var}\varphi^{(h)}(\vk,t) = \frac12 \,
\nu^{(h)}_{eL}(\vk, 0)\, \Bigg[\left(
          \begin{array}{c}
            1+\mathcal{C}^{(h)}_{-h}\\
             -\mathcal{S}^{ (h) }_{-h}  \\
          \end{array}\right)  \, e^{-i\omega^{ (h) }_1(k,-h)\,t} + \left(
          \begin{array}{c}
            1-\mathcal{C}^{ (h) }_{-h}\\
             \mathcal{S}^{(h)}_{-h}  \\
\end{array}\right)  \, e^{-i\omega^{ (h) }_2(k,-h)\,t} + \mathcal{O}\Big(
\frac{\overline{M}^{\,2}}{k^2}\Big) \Bigg],
          \eea

\bea \label{xi} \xi^{(h)}(\vk,t) & = &   \frac12
\,\nu^{(h)}_{eL}(\vk, 0)\, \left(\frac{h\,\overline{M}}{2 \;
k}\right) \,\,\Bigg\{\,\,\left(
          \begin{array}{c}
            1+\mathcal{C}^{(h)}_{-h}\\
             -\mathcal{S}^{ (h) }_{-h}  \\
          \end{array}\right)  \, e^{-i\omega^{ (h) }_1(k,-h)\,t} + \left(
          \begin{array}{c}
            1-\mathcal{C}^{ (h) }_{-h}\\
             \mathcal{S}^{(h)}_{-h}  \\
\end{array}\right)  \, e^{-i\omega^{ (h) }_2(k,-h)\,t}
\nonumber \\
&&
    - \left(
          \begin{array}{c}
            1+\cos 2\theta \\
             -\sin 2\theta   \\
          \end{array}\right)  \, e^{ -i\omega^{(h)}_1(k,h)\,t} -  \left(
          \begin{array}{c}
            1-\cos 2\theta \\
             \sin 2\theta  \\
          \end{array}\right)  \, e^{ -i\omega^{(h)}_2(k,h)\,t} +
\mathcal{O}\Big( \frac{\overline{M} }{k }\Big)
          \,\,\Bigg\},          \eea

\no where $\varphi^{(h)}(\vk,t)$ and $ \xi^{(h)}(\vk,t)$ are the
flavor doublets corresponding to the left-handed and right-handed
neutrinos with helicity $h$ respectively. The upper component
corresponds to the electron neutrino $\nu^{(h)}_e(\vk,t)$ while the
lower component corresponds to the muon neutrinos
$\nu^{(h)}_\mu(\vk,t)$. The factors $\mathcal{C}^{(h)}_{\lambda}(k)$
and $\mathcal{S}^{(h)}_{\lambda}(k)$ are defined as

\bea \mathcal{C}^{(h)}_{\lambda}(k) =
\cos\left[2\theta^{(h)}_m(\lambda k) \right]~;~~
\mathcal{S}^{(h)}_{\lambda}(k) = \sin\left[2\theta^{(h)}_m(\lambda
k) \right]\, . \eea

The suppression factor $M/k$ in the right handed component
(\ref{xi}) is of course a consequence of the chirality flip
transition from a mass term in the relativistic limit. For
relativistic neutrinos and more specifically for neutrinos in the
medium prior to BBN with $k\sim T \sim \mathrm{few~MeV}$ the right
handed component is negligible as expected.

The one-loop computation of the self-energy performed above does not
include absorptive processes such as collisions of neutrinos with
leptons (or hadrons) in the medium. Such absorptive part will emerge
in a two loops calculation and is of $\mathcal{O}(G^2_F)$. While we
have not calculated these contributions it is clear from the
analysis what it should be expected: the frequencies
$\omega_{1,2}(k)$ are the ``exact'' dispersion relations of the
single particle poles of the Dirac propagator in the medium. At two
loops the self energy will feature an imaginary part with support on
the mass shell of these single particle states. The imaginary part
of the self-energy evaluated at these single particle energies yield
the \emph{width} of the single quasi-particle states
$\Gamma_1(k);\Gamma_2(k)$ and the oscillatory exponentials in the
expressions above are replaced as follows

\be e^{-i\omega_a(k)\,t} \rightarrow e^{-\Gamma_a(k) \,t} \,
e^{-i\omega_a(k)\,t}~~;~~ a=1,2 \label{damping} \ee

While our one loop calculation does not include the damping rates
$\Gamma_a$ we will invoke results available in the
literature\cite{notzold,enqvist,barbieri,dolgov} to estimate the
collisional relaxation time scales (see section
\ref{sec:timescales}).

The corresponding fields for the left-handed and right-handed
component neutrinos in configuration space are obtained by
performing the spatial Fourier transform

\bea \label{varmp_fourier}\varphi^{(h)}(\vec{r},t )= \int \frac{d^3
\vk}{(2\pi)^3} \,\,\varphi^{(h)}(\vk,t )\,\, e^{i\vk \cdot \vec{r}},
\eea

\bea \label{xis_fourier}\xi^{(h)}(\vec{r},t )= \int \frac{d^3
\vk}{(2\pi)^3} \,\,\xi^{(h)}(\vk,t )\,\, e^{i\vk \cdot \vec{r}}.
\eea

For an arbitrary initial configuration these integrals must be done
numerically, but analytic progress can be made by assuming an
initial Gaussian profile, describing a wave-packet in momentum space
centered at a given momentum, $\vec{k}_0$ with a width $\sigma $.
While the width could generally depend on helicity we will consider
the simpler case in which it is the same for both helicities.
Namely, we consider

\be \nu^{(h)}_{eL}(\vk, 0)= \nu^{(h)}_{eL}( 0)\left(\frac{\pi}
{\si^2}\right)^{ \frac{3}{2}
}\,\exp\left[-\frac{(\vk-\vk_0)^2}{4\sigma^2}\right], \ee where
$\nu^{(h)}_{eL}( 0)$ is an arbitrary amplitude and assume that wave
packet is narrow in the sense that $\si \ll k_0$. In the limit
$\sigma  \rightarrow 0$ the above wave-packet becomes
$\nu^{(h)}_{eL}( 0)\delta^3(\vec{k}-\vec{k}_0)$. In the opposite
limit of large $\sigma $ the wave packet describes an inhomogeneous
distribution spatially localized within a distance $\approx 1/\sigma
$. For a narrow wave packet the momentum integral can be carried out
by expanding the integrand in a series expansion around $k_0$
keeping up to quadratic terms.

\subsection{Integrals}

The typical integrals are of the form

\be I(\vec{r},t)= \left(\frac{\pi} {\sigma^2}\right)^{ \frac{3}{2}
}\,\int \frac{d^3 k}{(2\pi)^3} \mathcal{A }(k) \,
\exp\left[-\frac{(\vk-\vk_0)^2}{4\sigma^2}+i\vk\cdot \vec{r}
-i\omega(k)\,t\right] \ee where $\mathcal{A}$ stands for the factors
$(1\pm\mathcal{C})\,;\mathcal{S}$ in eqns. (\ref{var},\ref{xi}), and
$\omega(k)$ are the general dispersion relations obtained above. The
computation of these integrals is simplified by noticing that for
any function $F(k)$ the expansion around $\vec{k}_0$ up to quadratic
order is given by

\be F(k)= F(k_0)+ F'(k_0)\,\widehat{\bf k}_0\cdot({\bf k}-{\bf
k}_0)+ \frac{1}{2}\left(F''(k_0)\,P^{\|}_{ij}(\widehat{\bf
k}_0)+\frac{F'(k_0)}{k_0}\,P^{\perp}_{ij}(\widehat{\bf
k}_0)\right)({\bf k}-{ \bf k}_0)_i({\bf k}-{\bf k}_0)_j +\cdots \ee
where \be P^{\|}_{ij}(\widehat{\bf k})= \widehat{\bf k}_{
i}\widehat{\bf k}_{ j} ~~;~~P^{\perp}_{ij}(\widehat{\bf k} ) =
\delta_{ij}-\widehat{\bf k}_{ i}\widehat{\bf k}_{ j} \ee

The result of the integration can be written more compactly by
introducing the following quantities

\bea\label{sigmast} \sigma^2_{\parallel}(t) & = & \sigma^2 \,
\frac{\left(1-i\,\frac{t}{\tau_{\parallel}}\right)}{\left[1
+\frac{t^2}{\tau^2_{\parallel}} \right]}
 \equiv \Phi_{\parallel}(t)\left(1-i\,\frac{t}{\tau_{\parallel}}\right) \\
\sigma^2_{\perp}(t) & = & \sigma^2 \,
\frac{\left(1-i\,\frac{t}{\tau_{\perp}}\right)}{\left[1
+\frac{t^2}{\tau^2_{\perp}} \right]} \equiv
\Phi_{\perp}(t)\left(1-i\,\frac{t}{\tau_{\perp}}\right)\nonumber
\eea where we have introduced the   perpendicular and parallel
dispersion time scales given respectively by

\be \label{taus}\tau_{\perp} = \frac{k_0}{2\sigma^2 v_g}~~;~~
\tau_{\parallel} = \frac{1}{2\sigma^2 \omega''(k_0)} = \gamma^2
\tau_{\perp}\,. \ee It will be seen in detail below that these two
time scales are indeed associated with the spreading of the wave
packet in the transverse and longitudinal directions.

The group velocity $v_g$ and effective Lorentz factor\footnote{For
the usual dispersion relation $\omega(k)=\sqrt{k^2+M^2}$ it is
straightforward to confirm that $\gamma^2=(1-v^2_g)^{-1}$} $\gamma$
are given by

\be \vec{v}_g= \omega'(k_0)\,\widehat{\bf{k}}_0 ~~;~~ \gamma^2 =
\frac{v_g}{k_0 \omega''(k_0)} \ee

The transverse and longitudinal coordinates are \be
\vec{X}_{\parallel}(t) = \widehat{\bf{k}}_0 \left( \vec{r}\cdot
\widehat{\bf{k}}_0- v_g t\right)~~;~~ \vec{X}_{\perp}= \vec{r}-
\widehat{\bf{k}}_0 \, \left(   \vec{r}\cdot \widehat{\bf{k}}_0
\right) \ee  and in terms of these variables we finally find

\be I(\vec{r},t) = \left[
\frac{\sigma_{\parallel}(t)\sigma^2_{\perp}(t)}{\sigma^3}\right]\,\mathcal{A}(k_0;\vec{r},t)\,
\,e^{i\left(\vec{k}_0\cdot\vec{r}- \Psi( \vec{r},t)\,t\right)}\,
 \, e^{-
\left(\Phi_{\perp}(t)\vec{X}^2_{\perp}+ \Phi_{\parallel}(t)
\vec{X}^2_{\parallel}(t) \right)} \label{integral} \ee where the
phase

\be   \Psi( \vec{r},t) = \omega(k_0) +
\frac{\Phi_{\perp}(t)}{\tau_{\perp}}\,X^2_{\perp}+
\frac{\Phi_{\parallel}(t)}{\tau_{\parallel}}\,X^2_{\parallel}(t)\label{phaseofrt}\ee
and \be \mathcal{A}(k_0;\vec{r},t) = \mathcal{A}(k_0 )+ 2\, i\,
\mathcal{A}'(k_0 )\sigma^2_{\parallel}(t)\,\widehat{\bf{k}}_0\cdot
\vec{X}_{\parallel}(t)+ \mathcal{A}''(k_0 )
\sigma^2_{\perp}(t)\,\left(1-2\sigma^2_{\perp}(t)
\vec{X}^2_{\perp}\right)+ \frac{\mathcal{A}'(k_0
)}{k_0}\sigma^2_{\parallel}(t)\left(1-2\sigma^2_{ \parallel}(t)\,
\vec{X}^2_{\parallel}(t)\right) \ee

Neglecting the prefactor $\mathcal{A}(k_0;\vec{r},t)$ we see that

\be \label{ampli} |I (\vec{r},t)|^2 \propto
\Bigg[\Big(1+\frac{t^2}{\tau^2_\parallel}\Big)\Big(1+\frac{t^2}{\tau^2_\perp}\Big)^2
\Bigg]^{-\frac12}\, e^{- 2\left(\Phi_{\perp}(t)\vec{X}^2_{\perp}+
\Phi_{\parallel}(t) \vec{X}^2_{\parallel}(t) \right)} \ee describes
a wave-packet moving in the direction parallel to the momentum
$\vec{k}_0$ with the group velocity $v_g$ and dispersing both along
the perpendicular and parallel directions. The expressions for
$\Phi_{\perp}(t)$ and $\Phi_{\parallel}(t)$ given by eqn.
(\ref{sigmast}) clearly show that the dispersion time scales along
the parallel direction and transverse directions are given by
$\tau_{\parallel},\tau_{\perp}$ respectively and $\tau_{\parallel}$
displays the time dilation factor $\gamma$. The wave packet is
localized in space within a distance of order
$1/\sqrt{\Phi(t)}\propto 1/\sqrt{\sigma}$ in either direction. Small
$\sigma$ localizes the wave packet in momentum space while large
$\sigma$   the wave packet is spatially localized. For large
$\sigma$ the integrals must necessarily be performed numerically.

This discussion highlights that the derivative terms in the
prefactor $\mathcal{A}(k_0;\vec{r},t)$, which are a consequence of
the \emph{momentum dependence of the mixing angles} correspond to an
expansion in the ratio $\sigma/k_0$. This can be understood from the
following argument: $\mathcal{A} \sim
(1\pm\mathcal{C}),\mathcal{S}$, hence its derivatives with respect
to momentum are of the form $ f(k) \Delta'$ with $f(k)$ being smooth
and bounded functions of $\mathcal{O}(1)$, while $\Delta$ is at most
of the form $\Delta_0 k + \Delta_1 k^2$ in the relativistic limit,
(see eqn. (\ref{Deltah})) therefore $\Delta' \approx \Delta/k$.
These derivatives multiply powers of $\sigma_{\perp,\parallel}
X_{\perp,\parallel}$, and the exponential damping in $I$  restricts
these contributions to the range $|\sigma_{\perp,\parallel}
X_{\perp,\parallel}|
 \approx 1$. Therefore in the narrow packet approximation $\sigma
 \ll k_0$ the higher order derivative terms are suppressed by
 powers of $\sigma/k_0 \ll 1$. We have invoked this narrow packet approximation
 to carry out the momentum integral, therefore consistently with
 this approximation we will only keep the first derivative term, which is of
 $\mathcal{O}(\sigma/k_0)$ and neglect the higher order
 derivatives, which are of higher order in this ratio. Namely in
 the analysis that follows we approximate

 \be |\mathcal{A}(k_0;\vec{r},t)|^2 \approx  |\mathcal{A}(k_0 )|^2\left[1+
 4 \, \frac{\mathcal{A}'(k_0 )}{\mathcal{A}(k_0
)}\Phi_{\parallel}(t)\,\widehat{\bf{k}}_0\cdot
\vec{X}_{\parallel}(t)\,\frac{t}{\tau_{\parallel}}\right]
\label{Aaprox}\ee

  In this manner we consistently keep the lowest order
 corrections arising from the \emph{momentum dependence of the mixing
 angles in the medium}.

 We now have all the ingredients for our analysis of the space
 time evolution. The above general expressions for the time
 evolution of initially prepared wave-packets, eqns.
 (\ref{var},\ref{xi}) combined with the dispersion relations
 obtained in section (\ref{disprel}) provide a solution to the most
 general case. We focus our discussion on the case of the early
 Universe, in which the typical
 neutrino energies are $\sim \textrm{MeV}$. With (active) neutrino
 masses in the range $M_a \sim \textrm{eV}$ and $\delta M^2 \sim
 10^{-5}-10^{-3}$ it is clear from the results above that the
 amplitude of the right handed component is suppressed by a factor
 $\overline{M}/k \sim 10^{-6}$ and the medium corrections to the
 dispersion relations for positive energy neutrinos with positive
 helicity  and negative energy neutrinos with negative helicity are
 suppressed by a factor $\overline{M}^{\,2}/k^2$ with respect to the
 opposite helicity assignement. Therefore in what follows we focus
 our discussion to the case of left handed negative helicity
 neutrinos (and positive helicity antineutrinos).

 \subsection{Space-time evolution and oscillations.}

We now focus on describing the evolution of negative helicity
neutrinos or positive helicity antineutrinos.

 The initial state considered above corresponds to a wave-packet of
 electron neutrinos at $t=0$ but no muon neutrinos. The lower
 component of the flavor spinor in eqn. (\ref{var},\ref{varmp_fourier})
describes the
 wave-packet of the muon neutrino at any arbitrary time.
 We begin by studying the transition probability from an initial
electron neutrino wave packet of negative helicity to a muon
neutrino wave packet.

 Using the results obtained
 in the previous section for the integrals in the
 narrow packet approximation we find the transition probability

\bea \label{transpro}  \mathcal{P}_{e\rightarrow \mu}(\vec{r},t) =
|\nu^{(h)}_{\mu\,L}(\vec{r}, t)|^2 & =  & \frac14 \,
|\nu^{(h)}_{eL}(\vk, 0)|^2 |\mathcal{S}(k_0 )|^2\left[1+
 4 \, \frac{\mathcal{S}'(k_0 )}{\mathcal{S}(k_0
)}\overline{\Phi}_{\parallel}(t)\,\widehat{\bf{k}}_0\cdot
\vec{X}_{\parallel}(t)\,\frac{t}{\overline{\tau}_{\parallel}}\right]\times
\nonumber
\\ & &
\Bigg[|I_1(\vec{r},t)|^2+|I_2(\vec{r},t)|^2-2\,|I_1(\vec{r},t)|\,|I_2(\vec{r},t)|\,\cos\left[
\left(\Psi_1(\vec{r},t)-\Psi_2(\vec{r},t)\right)\,t\right]
  \Bigg] \eea where $\mathcal{S}= \sin\left[2\theta^{(h)}_m(\pm
k) \right]$ and $I_{1,2}(\vec{r},t)~;~\Psi_{1,2}$ correspond to the
integrals  and phases given by eqn. (\ref{integral},\ref{phaseofrt})
with  the frequencies $\omega_{1,2}(k)$ for negative helicity given
by eqns. (\ref{posenegh}).   In the expression above we have taken a
common \emph{prefactor} by neglecting the differences between the
group velocities and the masses, taking $v_g =1$,  and
$\overline{\Phi}_{\parallel},\overline{\tau}_{\parallel}$ correspond
to $\Phi_{\parallel},\tau_{\parallel}$ with a mass $\overline{M}$.
We focus our attention on the interference term which is the
space-time manifestation of the oscillation phenomenon and features
the oscillatory cosine function. The amplitude of the oscillation,
$|I_1(\vec{r},t)I_2(\vec{r},t)|$ describes the product of two
wavepackets of the form given by eqn. (\ref{ampli}).

It is convenient to write the product $|I_1\,I_2|$ in the following
form

\be \label{overlap} |I_1(\vec{r},t)\,I_2(\vec{r},t)| \approx
\Bigg[\Big(1+\frac{t^2}{\tau^2_\parallel}\Big)\Big(1+\frac{t^2}{\tau^2_\perp}\Big)^2
\Bigg]^{-\frac{1}{2}}\, e^{-  \left(\Phi_{\perp,1}(t) +
\Phi_{\perp,2}(t) \right)\vec{X}^2_{\perp }}\, e^{- \Phi_{ CM}(t)
\vec{X}^2_{ CM}(t) } \, e^{-
 \Phi_{ R}(t)    X^2_{ R}(t)}  \ee where we have
 introduced the center of mass (CM) and relative (R) variables

 \bea \vec{X}_{ CM} & = & \widehat{\bf{k}}_0 \left( \vec{r}\cdot
\widehat{\bf{k}}_0- v_{CM}(t)\, t\right)~~;~~v_{CM}(t)=
\frac{\Phi_{\parallel\,1}(t)\, v_{g\,1} +\Phi_{\parallel\,2}(t)\,v_{
g\,2}}{\Phi_{\parallel\,1}(t)+\Phi_{\parallel\,2}(t)}\\\vec{X}_{ R}
& = & \vec{X}_{
\parallel\,1}-\vec{X}_{ \parallel\,2}=-
\left(\vec{v}_{g\,1}-\vec{v}_{g\,2}\right)t \\
 \Phi_{ CM} & = & \Phi_{\parallel\,1}+\Phi_{\parallel\,2}
 ~~;~~ \Phi_{ R}   =
\frac{\Phi_{\parallel\,1}\,\Phi_{\parallel\,2}}{\Phi_{\parallel\,1}+\Phi_{\parallel\,2}}
 \eea

The   integral (\ref{overlap}) describes the product of two gaussian
wave packets spreading in the transverse and longitudinal directions
and separating in the longitudinal direction because of the
difference in group velocities, made explicit by the term $\Phi_{
R}(t) X^2_{ R}(t)$.

The first two terms in eqn. (\ref{transpro}) describe the incoherent
sum of the probabilities associated with separated wave packets of
propagating mode  eigenstates , in the third, interference term, the
product $|I_1||I_2|$ is the overlap between these two wave-packets
that are slowly separating because of different group velocities. As
discussed above a two loop calculation of the self-energies will
lead to a quasiparticle \emph{width} and a damping rate $\Gamma_a$
for the individual quasiparticle modes of frequency $\omega_a(k)$,
the discussion leading up to eqn. (\ref{damping}) suggests that the
integrals

\be |I_a(\vec{r},t)| \rightarrow
e^{-\Gamma_a(k)\,t}\,|I_a(\vec{r},t)|\,. \ee

\subsubsection{  Coherence and ``freeze-out'' }

 Since $\vec{X}_R = (\vec{v}_{g\,2}-\vec{v}_{g\,1})t$
does not depend on position, the overlap between the separating wave
packets becomes vanishingly small for $t>>t_{coh}$ where the
coherence time scale $t_{coh}$ is defined by

\be\label{cohtime} \Phi_{ R}(t_{coh})
(\vec{v}_{g\,2}-\vec{v}_{g\,1})^2\,t^2_{coh} =1\ee

Before we engage in an analysis of the different cases, it is
important to recognize that there are several dimensionless small
ratios: i) $\sigma/k_0 \ll 1$ describes narrow wave-packets, this
approximation was implemented in the calculation of the integrals,
ii) $\overline{M}/k \ll 1$ in the relativistic limit with $k \sim
\textrm{MeV}$ for example in the early Universe near the epoch of
BBN or for supernovae, iii) $\delta M^2 / \overline{M}^{\,2} \ll 1$
describes a nearly degenerate hierarchy of neutrino masses. Since in
the relativistic limit $v_{1\,g}-v_{2\,g} \sim \delta M^2/k^2$ we
can neglect the difference in the masses in $\Phi_{\parallel}$ and
write $\Phi_{\parallel \,1} \sim \Phi_{\parallel \,2} \sim
\overline{\Phi}_{\parallel}$ where the masses are replaced by the
mean mass $\overline{M}$ given by eqn. (\ref{mbardm}), and similarly
for $\Phi_{\perp}$. Therefore to leading order in small quantities
we can replace $\Phi_R$ above by $\overline{\Phi}_{\parallel}/2$
leading to \be \Phi_{CM}(t) = 2 \, \overline{\Phi}_{\parallel}(t) =
\frac{2\,\sigma^2}{1+\frac{t^2}{\overline{\tau}^2_{\parallel}}}
~~;~~ v_{CM}= \frac{1}{2} (v_{g\,1}+v_{g\,2}) \label{CMVars}\ee
where $\overline{\tau}_{\parallel}$ is given by eqn. (\ref{taus})
for $\overline{M}$, and \be \label{overlap2}
 \frac{1}{2}\overline{\Phi}_{\parallel}(t)    X^2_{ R}(t)   =
\frac{\left(\frac{t}{t_c}\right)^2}{1+\left(\frac{t}{\overline{\tau}_{\parallel}}\right)^2}
  \ee where we have introduced the time scale

  \be \label{tc} t_c = \frac{\sqrt{2}}{\sigma |{v}_{g\,2}-
  {v}_{g\,1}|} \ee

The coherence time scale is the solution of the equation

\be\label{cohtime}
\frac{\left(\frac{t_{coh}}{t_c}\right)^2}{1+\left(\frac{t_{coh}}{\overline{\tau}^{\,2}_{\parallel}}\right)^2}
= 1 \ee

The expression (\ref{overlap2}) reveals a remarkable feature: for
$t\gg \overline{\tau}_{\parallel}$ the overlap between the
separating wave-packets saturates to a \emph{time independent value}

\be \label{freeze} \frac{1}{2}\overline{\Phi}_{\parallel}(t)  X^2_{
R}(t) \rightarrow
\left(\frac{\overline{\tau}_{\parallel}}{t_c}\right)^2 \,. \ee  This
effect
 has been recognized in ref.\cite{Beuthe} and results from the
 longitudinal dispersion catching up with the separation of the wave
 packets. This phenomenon is relevant only in the case when $t_{c}
 > \overline{\tau}_{\parallel}$ in which case the overlap of the separating
 wave packets ``freezes'' and the packets maintain coherence for the
 remainder of their evolution.  There are two distinct
 possibilities:

 \bea \label{casea}     && t_{c}\ll \overline{\tau}_{\parallel}
~~~~\textbf{:~(a)} \label{casea} \\
 && t_{c}\gg \overline{\tau}_{\parallel} ~~~~\textbf{:~(b)} \eea      In case
{\bf (a)}  we can
 approximate \be \frac{1}{2}\overline{\Phi}_{\parallel}(t)    X^2_{ R}(t)
 \approx \left( \frac{t}{t_c}\right)^2 \ee since during the time interval in
which the
 separating packets maintain coherence $t \ll t_c \ll
\overline{\tau}_{\parallel}$  and in this case the relevant
coherence time scale is
 $t_c$.

 In case {\bf (b)}  the ``freeze out'' of  coherence results
 and the long time limit of the overlap between the wave packets in
 the longitudinal direction remains large and determined by eqn.
 (\ref{freeze}).

However, while this ``freezing of coherence'' phenomenon in the
longitudinal direction ensues in this regime, by the time when the
coherence freezes $t \sim \overline{\tau}_{\parallel}$ the wave
packet has spread dramatically in the \emph{transverse} direction.
This is because of the enormous Lorentz time dilation factor in the
longitudinal direction which ensures that $t \sim
\overline{\tau}_{\parallel} \gg \tau_{\perp}$ (see eqn.
(\ref{taus})). The large spreading in the transverse direction
entails a large suppression of the transition probability

\be \mathcal{P}_{e\rightarrow \mu}(\vec{r},t\sim
\overline{\tau}_{\parallel}) \propto
\left(\frac{\tau_{\perp}}{\overline{\tau}_{\parallel}}\right)^2 \sim
\frac{1}{ \gamma ^4}\sim \left(\frac{\overline{M}}{k_0}\right)^4 \,
.\ee

For $\overline{M} \sim \mathrm{eV}$ and $k_0 \sim \mathrm{MeV}$ the
above ratio is negligible. Therefore while the phenomenon of
freezing of coherence is remarkable and fundamentally interesting,
it may not lead to important consequences because the transition
probability is strongly suppressed in this regime. Therefore in the
time scale during which the transition probability is
non-negligible, namely $t \ll \overline{\tau}_{\parallel}$  the
overlap integral can be simplified to \be
e^{-\frac{1}{2}\overline{\Phi}_{\parallel}(t) X^2_{ R}(t) } \approx
e^{- \left( \frac{t}{t_c}\right)^2 }\,. \ee

\subsubsection{Effective oscillation frequency}

 Another aspect of the interference term is the \emph{effective time
 dependent oscillation frequency} $ \Psi_1(\vec{r},t) - \Psi_2(\vec{r},t)
 $ where  the $\Psi_{a}$ are given by eqn. (\ref{phaseofrt}) for the
 frequencies $\omega_a(k)$ of the propagating modes. The
 spatio-temporal dependence of this effective phase is a consequence
 of the dispersion of the inhomogeneous configurations, encoded in
 the functions $\Phi$ and results in a \emph{drift} of the
 oscillation frequency, a result that confirms a similar finding in the vacuum
case in ref.\cite{bernardini}.
  Because of the exponential fall off of the
 amplitudes the maximum value of the \emph{drift} contribution is
 achieved for $\Phi_{\perp,a}X^2_{\perp} \sim
 1~~;~~\Phi_{\parallel,a}X^2_{\parallel,a}(t)\sim 1$, namely in front and back
of the center of the wave-packets, both
 in the transverse and the longitudinal directions. Furthermore,
 because of the Lorentz dilation factor, $\tau_{\parallel} \gg
 \tau_{\perp}$ for relativistic neutrinos. Therefore we can
 approximate the effective oscillation frequencies as

 \be \Psi_1 -\Psi_2 \sim \omega_1(k)-\omega_2(k)  +
\frac{2\sigma^2}{k_0}(v_{g,1}-v_{g,2})\label{effphase} \ee

The dispersion relations and mixing angles obtained above along with
the the results (\ref{var},\ref{xi}),
   yield the  complete space-time evolution for wave-packets with
   initial conditions corresponding to an electron neutrino. Rather than
studying the general case, we focus on three different situations
which summarize the most general cases,i) $\Delta_h/\delta M^2 \ll
1$ corresponding to the case of vacuum oscillations, ii)
$\Delta_h/\delta M^2 \sim \cos2\theta$ corresponding to a resonance
in the medium, and iii) $\Delta_h/\delta M^2 \gg 1$ corresponding to
the case a hot and or dense medium in which oscillations are
suppressed.

\subsection{$ \Delta_h/\delta M^2 \ll 1$: vacuum
oscillations.}

We study   this case  not only to compare to results available in
the literature, but also to establish a ``benchmark'' to compare
with the results with medium modifications. Beuthe \cite{Beuthe} has
studied the propagation of neutrino wave-packets in the vacuum case
including dispersion and in ref.\cite{bernardini} an effective
frequency similar to (\ref{phaseofrt}) has been found for
wave-packets propagating in the vacuum. In this case for positive
energy, negative helicity neutrinos with $a=1,2$

\be \omega_{a}(k_0) \sim k_0 + \frac{M^2_{a}}{2k_0} ~~;~~ v_{g,a}
\sim 1-\frac{M^2_{a}}{2k^2_0}~~;~~\omega''_{a}(k_0) \sim
\frac{M^2_{a}}{ k^3_0} \ee leading to the vacuum time scales \bea
t_{c,v}   & = &  \frac{2\sqrt{2}k^2_0}{\sigma
|\delta\,M^2|}\label{tcv}\\
\overline{\tau}_{\parallel} & = &
\frac{k^3_0}{2\sigma^2\overline{M}^{\,2}} \label{tparav} \eea


In the case when  \be
\left|\frac{k_0}{\sigma}\,\frac{\delta\,M^2}{4\,\overline{M}^{\,2}}
\right| \gg 1 \ee   the vacuum  coherence time is given by \be
t_{c,v} = \overline{\tau}_{\parallel}
\left|\frac{\sigma}{k_0}\,\frac{4\,\overline{M}^{\,2}}{\delta\,M^2}
\right| = \frac{2k^2_0}{\sigma |\delta\,M^2|}\ll
\overline{\tau}_{\parallel}
 \, \ee and the overlap between
the separating wave-packets vanishes well before the packets
disperse appreciably along the longitudinal direction. On the other
hand, in the case when \be
\left|\frac{k_0}{\sigma}\,\frac{\delta\,M^2}{4\,\overline{M}^{\,2}}
\right| \ll 1 \ee the spreading of the wave packets catches up with
the separation and the overlap between them freezes when $t \equiv
t_{f,v} = \overline{\tau}_{\parallel} = k^3_0/2\sigma^2
\overline{M}^{\,2}$. With $|\delta M^2|/4\overline{M}^{\,2}\sim
10^{-4}$ for solar or $\sim 10^{-3}$ for atmospheric neutrinos the
phenomenon of ``freezing'' of the overlap and the survival of
coherence is available for \be 1 \ll \frac{k_0}{\sigma} \ll
\frac{4\,\overline{M}^{\,2}}{|\delta\,M^2|}\ee which is well within
the ``narrow wave packet'' regime. However, as discussed above, when
the coherence freezes the transition probability has been strongly
suppressed by transverse dispersion. Therefore during the time scale
during which the transition probability is non-negligible we can
approximate the exponent in (\ref{overlap})

\be  \frac{1}{2}\overline{\Phi}_{\parallel}(t)    X^2_{ R}(t) \sim
  \left(\frac{t}{t_{c,v}}\right)^2 \ee

The effective oscillation frequency is given by eqn.
(\ref{effphase}) which becomes

\be \Psi_1-\Psi_2 \sim \frac{\delta
M^2}{2k}\left(1-\frac{2\sigma^2}{k^2_0} \right)
\label{efffreqvac}\ee while the corrections tend to diminish the
oscillation frequency,   these   are rather small in the narrow
packet approximation.

\subsection{Medium effects: near resonance}

In refs.\cite{barbieri,enqvist,notzold,dolgov,hoboya} it was
established that if the lepton asymmetries are of the order of the
baryon asymmetry $\eta \sim 10^{-9}$ there is the possibility of a
resonance  for the temperature range $m_e \ll T \ll m_{\mu}$ for
positive energy negative helicity neutrinos with $\omega(k)  \sim
k+\overline{M}^2/2k~~;~~h=-1$ or positive energy positive helicity
antineutrinos with $\omega(k) \sim -k-\overline{M}^2/2k~~;~~h= 1$
respectively. It is convenient  to introduce the following notation
\bea \mathcal{L}_9 & = & 10^9
\left(\mathcal{L}_e-\mathcal{L}_\mu\right)
\label{L9} \\
\delta_5 & = & 10^5 \left( \frac{\delta M^2}{\mathrm{eV}^2}\right)
\label{del5} \eea \emph{If} the lepton and neutrino asymmetries are
of the same order of the baryon asymmetry, then $ 0.2 \lesssim
|\mathcal{L}_9| \lesssim 0.7 $ and the fitting from solar and
KamLAND data suggests $|\delta_5 | \approx 8$. In this temperature
regime we find\cite{hoboya} for positive energy, negative helicity
neutrinos

\be \label{case1I} \frac{\Delta_{-}(k,k)}{\delta M^2} \approx
\frac{4}{\delta_5}\left(\frac{0.1\,T}{\mathrm{MeV}} \right)^4 \;
\frac{k}{T} \;  \Bigg[-\mathcal{L}_9 +
\left(\frac{2\,T}{\mathrm{MeV}} \right)^2\; \frac{k}{T} \Bigg] \; .
\ee and for positive energy positive helicity antineutrinos

\be \frac{\Delta_{+}(-k,k)}{\delta M^2} \approx
\frac{4}{\delta_5}\left(\frac{0.1\,T}{\mathrm{MeV}} \right)^4\;
\frac{k}{T} \Bigg[ \mathcal{L}_9 + \left(\frac{2\,T}{\mathrm{MeV}}
\right)^2 \; \frac{k}{T} \Bigg] \; . \ee

In the above expressions we have neglected terms of order $
\frac{\overline{M}^{\,2}}{k^2} $. With $k \sim T$ and in the
temperature regime just prior to BBN with $T \sim
\textrm{few}\,\textrm{MeV}$ the lepton asymmetry contribution
$\mathcal{L}$ is much smaller than the momentum dependent
contribution and will be neglected in the analysis that follows,
therefore  we  refer  to $\Delta_{h}(\lambda k,k)$ and
$S_{h}(\lambda k,k)$ as $\Delta(k)$ and $S(k)$ respectively since
these are independent of $h,\lambda$ in this regime. In this
temperature regime  we find for both cases (negative helicity
neutrinos and positive helicity antineutrinos) the following simple
expressions

\be \Delta(k)  \approx \frac{56\pi^2}{45\,\sqrt{2}}\frac{G_F k^2
T^4}{M^2_W} ~~;~~ S(k) \approx \Delta(k)(1+\cos^2\theta_w)
\label{hiT}.\ee

A resonance is available when $\Delta(k_0) \sim \delta M^2
\cos2\theta$, which may occur in this temperature regime for $k_0
\sim T \sim 3.6
\,\textrm{MeV}$\cite{barbieri,enqvist,notzold,dolgov,hoboya} for
large mixing angle ($\theta_{12}$) or $k\sim T \sim 7\,\mathrm{MeV}$
for small mixing angle ($\theta_{13}$). Near the resonance the
in-medium dispersion relations and group velocities are given by

\bea   \omega_a(k) & \approx &  k+\frac{M^2_a}{2k} -\frac{\delta
M^2}{4 k }\Bigg\{(1+\cos^2\theta_w)\cos2\theta+(-1)^{a-1}\, (1-\sin2
\theta) \Bigg\}\label{dispermed} \\ v_{g,a}  & \approx &
1-\frac{M^2_a}{2k^2} +\frac{\delta M^2}{4 k^2
}\Bigg\{(1+\cos^2\theta_w)\cos2\theta+(-1)^{a-1}\, (1-\sin2 \theta)
\Bigg\}\label{groupvelmed}
 \eea

Again we focus our discussion on the interference terms in the
transition probability (\ref{transpro}), in particular the medium
modifications to the oscillation frequencies and coherence time
scales. To assess these   we note the following (primes stand for
derivatives with respect to $k$):

\be \Omega(k)|_{res} = \sin 2\theta ~~; \Omega^{'}(k)|_{res} = 0
~~;~~ \Omega^{''}(k)|_{res} = \frac{4}{k^2}
\frac{\cos^2\theta}{\sin2\theta}  \ee which when combined with
equation (\ref{posenegh}) yield

\be \omega_1(k)-\omega_2(k) = \frac{\delta M^2}{2k} \sin2\theta ~~;
~~ v_{g,1}-v_{g,2} \approx -\frac{\delta M^2}{2k^2}\sin2\theta
\label{omegdif} \ee

We also note that near the resonance \be \sin'2\theta_m(k) \propto
\cos2\theta_m \approx 0 \label{nocorr}\,, \ee therefore the
corrections arising from the energy dependence of the mixing angle
in the transition probability (\ref{transpro}) become
\emph{vanishingly small}. The transverse and longitudinal dispersion
time scales are given by

\bea \tau_{a\perp} = \frac{k_0 }{2\sigma^2\,v_{g,a}} ~;~~
\tau_{a\parallel} \app \overline{\tau}_{\parallel}\left[
1-(-1)^{a-1}\frac{\delta M^2}{2\OM^2}\frac{1+\cos^{2}2\theta}{\sin2
\theta}\right]~;~~
\overline{\tau}_{\parallel}=\frac{k_0^3}{2\sigma^2 \OM^2}. \eea

Therefore in the medium near the resonance, the argument of the
exponential that measures the overlap between the separating wave
packets is  given by

\be \frac{1}{2}\overline{\Phi}_{\parallel}(t)    X^2_{ R}(t) \sim
\left(\frac{k_0}{\sigma}\,\frac{\delta\,M^2}{4\,\overline{M}^{\,2}}
\,\sin 2\theta\right)^2 \frac{
{t^2}}{{\overline{\tau}^2_{\parallel}} + t^2 } = \frac{\left(
\frac{t}{t_{c,m}}\right)^2}{1+
\left(\frac{t}{\overline{\tau}_{\parallel}} \right)^2 } \ee where
\be t_{c,m} = \frac{2 k^2_0}{\sigma |\delta\,M^2|\sin2\theta} =
\frac{t_{c,v}}{\sin2\theta} \label{cohtimes}\,. \ee

The effective oscillation frequency (\ref{effphase}) is given by

\be  \Psi_1-\Psi_2 \sim \frac{\delta M^2}{2k}\sin2\theta
\left(1-\frac{2\sigma^2}{k^2_0} \right) \label{efffreqmed}\ee which
when compared to the vacuum result (\ref{efffreqvac}) confirms the
relation between the vacuum and in-medium oscillation time scales
(\ref{reltimes}) since near the resonance $\sin2\theta_m \sim 1$.

We conclude that the main effects from the medium near the resonance
are an increase in the coherence   and in the oscillation time scale
$T = 2\pi/|\Psi_1-\Psi_2|$ by a factor $1/\sin2\theta$. For solar
neutrino mixing with $\sin2\theta_{12}\sim 0.9$ the increase in
these time scales is at best a  $10\%$ effect, but it becomes much
more pronounced in the case of atmospheric neutrino mixing since
$\sin2\theta_{13} \ll 1$.

\vspace{1cm}

\subsubsection{$\Delta_h /\delta M^2 \gg 1$ oscillation suppression by
the medium.}

In the temperature or momentum regime for which $\Delta_h  /\delta
M^2 \gg 1$ the expression for the in-medium mixing angles
(\ref{SinCos}) reveals that $\cos2\theta_m \rightarrow -1$. In this
case the in-medium mixing angle reaches $\theta_m \rightarrow \pi/2$
and the transition probability $\mathcal{P}_{e\rightarrow \mu}$
vanishes. Eqn. (\ref{var}) shows that in this case an electron
neutrino wavepacket of negative helicity propagates as an eigenstate
of the effective Dirac Hamiltonian in the medium with a dispersion
relation \be \omega_2(k) \sim vk +\frac{M^2_2}{2k} ~~;~~ v = \left[
1-\frac{14}{45 \,\sqrt{2} }\frac{G_F
T^4}{M^2_W}\left(1+\mathrm{sign}(\delta M^2)+\cos^2\theta_w
\right)\right]\ee where we have used eqn. (\ref{hiT}) for the case
when the momentum dependent contribution is much larger than the
asymmetries. The in-medium correction to the group velocity being
proportional to \be \frac{G_F T^4}{M^2_W} \sim 10^{-21}
\left(\frac{T}{\mathrm{MeV}}\right)^4 \ee is negligible in the
temperature regime in which the calculation is reliable, namely for
$T \ll M_{W}$.

\section{Time scales in the resonance
regime}\label{sec:timescales}

There are several important time scales that impact on the dynamics
of wave-packets in the medium as revealed by the discussions above,
but also there are two more relevant time scales that are pertinent
to a plasma  in an expanding Universe: the Hubble time scale $t_H
\sim 1/H$ which is   the cooling time scale $T(t)/\dot{T}(t)$ and
the collisional relaxation time scale $t_{rel}=1/\Gamma$ with
$\Gamma$ the weak interaction collision rate. Neither $t_H$ nor
$t_{rel}$  has been input explicitly in the calculations above which
assumed a medium in equilibrium and considered self-energy
corrections only up to $\mathcal{O}(G_F)$. The damping factor that
leads to the decoherence from neutral and charged current
interactions has been studied in detail in
references\cite{barbieri,enqvist,notzold,dolgov} and we take this
input from these references in order to compare this time scale for
damping and decoherence to the time scales for the space-time
evolution of the wave packets obtained above at the one-loop level.

In the temperature regime $1\,\mathrm{MeV} \leq T \leq
100\,\mathrm{MeV}$ the Hubble time scale is\cite{Turner} \be t_H
\sim   0.6 \left(\frac{T}{\mathrm{MeV}}\right)^{-2}\,s \ee and the
collisional rate is estimated to
be\cite{barbieri,enqvist,notzold,dolgov}

\be \Gamma \sim 0.25\, G^2_F T^5 \sim 0.25 \times 10^{-22}
\left(\frac{T}{\textrm{MeV}}\right)^5\,\textrm{MeV} \Rightarrow
t_{rel}   \sim 1.6 \left(\frac{T}{\textrm{MeV}}\right)^{-5}\, s \ee

In order to determine  the relevant time scales an estimate of the
momentum spread of the initial wavepacket $\sigma$ is needed. For
example, for neutrinos in the LSND experiment, the momentum spread
of the stopped muon is estimated to be about $0.01 MeV$
\cite{Stockinger}. An estimate of the momentum spread in the medium
can be the inverse of the mean-free path of the charged lepton
associated with the neutrino\cite{book1}. This mean free path is
determined by the electromagnetic interaction, in particular large
angle scattering, which can be simply estimated from one-photon
exchange to be $\lambda_{mf} \sim (\alpha^2_{em} T)^{-1}$. This
estimate yields \be \sigma \sim \alpha^2_{em} T \sim 10^{-4}
\left(\frac{T}{\mathrm{MeV}}\right)\, (\mathrm{MeV})\,. \ee For
neutrinos in the neutrinosphere of a core-collapse supernovae, the
estimate for $\sigma$ is also $\sim
10^{-2}\,\textrm{MeV}$\cite{book1}. We will take a value $\sigma
\sim 10^{-3}\,\mathrm{MeV}$ in the middle of this range as
representative to obtain order of magnitude estimates for the time
scales, but it is straightforward to modify the estimates if
alternative values of $\sigma$ can be reliably established.

We now consider the large mixing angle (LMA) case to provide an
estimate of the different time scales, but a similar analysis holds
for the case of small vacuum mixing (SMA) by an appropriate change
of $k_0;T$. Taking $k_0 \sim T \sim 3.6 \, \textrm{MeV},
\,\sigma\sim 10^{-3} \textrm{MeV}, \,|\delta M^2| \sim 8 \times
10^{-5} (\textrm{eV})^2, \,\OM\sim 0.25 eV$
 we obtain the following  time scales near the resonance region:

 i) {\bf oscillation time
 scales:}  \be   T_{vac} = \frac{4\pi k_0}{|\delta M^2|} \sim  3.8\times
10^{-4}\,s
 ~~;~~ T_{med} = \frac{T_{vac}}{\sin2\theta} \ee
ii) {\bf dispersion and coherence time scales:} \bea && \tau_{\perp}
\sim \frac{k_0}{2\sigma^2} \sim 1.2 \times 10^{-15}\,s
~~;~~\overline{\tau}_{\parallel} \sim \frac{k^3_0}{2\sigma^2
\overline{M}^{\,2}} \sim 0.25 \, s \\ && t_{c,v} =
\frac{2k^2_0}{\sigma|\delta M^2|} \sim 0.21 \,s ~~;~~ t_{c,m} =
\frac{t_{c,v}}{\sin 2\theta} \eea iii) {\bf expansion and
collisional relaxation time scales:} \be t_H \sim 4.6\times 10^{-2}
\,s ~~;~~ t_{rel} \sim 2.8\times 10^{-3}\,s \ee

For small vacuum mixing angle ($\theta_{13}$) the above results are
modified by taking $k_0\sim T \sim 7\,\mathrm{MeV}$.

In the resonance region the in-medium coherence time scale is of the
same order as the Hubble time (for LMA) or much longer (for SMA) and
and there is a large temperature variation during the coherence time
scale. However, the decoherence of the wave packets occurs on much
shorter time scales determined by the collisional relaxation scale
and the coherence time scale is not the relevant one in the medium
near the resonance.

Decreasing the momentum spread of the initial wave packet $\sigma$
\emph{increases} the dispersion and coherence time scales, with the
dispersion scales increasing faster. The medium effects are manifest
in an increase in the oscillation and the coherence time scales by a
factor $1/\sin2\theta$. This effect is more pronounced for $1-3$
mixing because of a much smaller mixing angle. It is clear from the
comparison between the coherence time scale in the medium $t_{c,m}$
and the relaxational (collisional) time scale $t_{rel}$ that unless
$\sigma$ is substantially \emph{larger} than the estimate above, by
at least one order of magnitude in the case of $1-2$  mixing, or
even more for $1-3$ mixing, collisions via neutral and charge
currents is the main source of  decoherence between the separating
wave-packets near the resonance. However, increasing $\sigma$ will
decrease the transverse dispersion time scale thus leading to
greater suppression of the amplitude of the wave packets through
dispersion.  Furthermore, for large mixing angle $\sin2\theta \sim
1$ the oscillation scale is \emph{shorter} than the collisional
decoherence time scale via the weak interactions $t_{rel}$,
therefore allowing several oscillations before the wave packets
decohere, and because the oscillation scale is much smaller than the
Hubble scale the evolution is adiabatic over the scale $t_{rel}$.
But for small mixing angle the opposite situation results and the
transition probability is suppressed by collisional decoherence,
furthermore for small enough mixing angle there is a breakdown of
adiabaticity. However, the strongest suppression of the survival
$\mathcal{P}_{e\rightarrow e}$ as well as the transition
$\mathcal{P}_{e\rightarrow \mu}$ probabilities (equally) is the
\emph{transverse} dispersion of the wave packets, on a time scale
many orders of magnitude \emph{shorter} than the decoherence
$t_{c,m}$ and the collisional $t_{rel}$ time scales. Unless
$\sigma^2$ is within the same order of magnitude of $|\delta M^2|$
the transverse dispersion occurs on time scales much faster than any
of the other relevant time scales and the amplitude of the
wave-packets is suppressed well before any oscillations or
decoherence by any other process can occur. Clearly a better
understanding of the initial momentum spread is necessary for a full
assessment of the oscillation probability in the medium.

\vspace{1cm}

\section{Discussions and Conclusions}\label{sec:DC}

In this article we implemented a non-equilibrium quantum field
theory method that allows to study the space-time propagation of
neutrino wave-packets directly from the effective Dirac equation in
the medium. The space-time evolution is studied as an initial value
problem with the full density matrix via linear response. The method
systematically allows to obtain the space-time evolution of left and
right handed neutrino wave packets.

A ``flavor neutrino'' wave packet evolves in time as a linear
superposition of wave-packets of ``exact'' (quasi) particle states
in the medium, described by the poles of the Dirac propagator
\emph{in the medium}. These states propagate in the medium with
different group velocities and the slow separation between these
packets causes their overlap to diminish leading to a loss of
spatial and temporal coherence. However, the time evolution of the
packets also features \emph{dispersion} as a result of the momentum
spread of the wave packets\cite{Beuthe}.

The space time dynamics feature a rich hierarchy of time scales that
depend on the initial momentum spread of the wave packet: the
transverse and longitudinal dispersion time scales $\tau_{\perp}\ll
\tau_{\parallel}$ which are widely separated by the enormous Lorentz
time dilation factor $\approx (k/\overline{M})^2$ with
$\overline{M}$ the average neutrino mass, and a coherence time scale
$t_{c,m}$ that determines   when the overlap of the wave packets
becomes negligible. The dynamics also displays the phenomenon of
``freezing of coherence'' which results from the competition between
the separation and   spreading of the wave packet along the
direction of motion (longitudinal). For time scales larger than
$\tau_{\parallel} $ the overlap of the wave-packets freezes, with a
large overlap in the case when $t_{c,m} \gg \tau_{\parallel}$, which
occurs for  a wide range of parameters.

We have focused on studying the space-time propagation in the
temperature and energy regime in which there is a resonance in the
mixing angle in the medium,   prior to
BBN\cite{barbieri,enqvist,notzold,hoboya}. Our main results are
summarized as follows:

\begin{itemize}
\item{Both the coherence and oscillation time scales are enhanced
in the medium with respect to the vacuum case by a factor
$1/\sin2\theta$ near the resonance, where $\theta$ is the vacuum
mixing angle. }

\item{There are small corrections to the oscillation formula from
the wave-packet treatment, but these are suppressed by two powers of
the ratio of the momentum spread  of the initial packet to the main
momentum.}

\item{There are also small corrections to the space-time evolution
from the energy dependence of the mixing angle, but these are
negligible near the resonance region.}

\item{The spreading of the wave-packet leads to the phenomenon of
``freezing of coherence'' which results from the competition between
the longitudinal dispersion and coherence time scales. This
phenomenon is a result of the longitudinal spreading of the
wave-packets ``catching up'' with their separation. Substantial
coherence remains frozen for $t_{c,m} \gg \tau_{\parallel}$. }

\item{We have compared the wide range of time scales present in
the early Universe when the resonance is available for $T\sim 3.6
\mathrm{MeV}$\cite{enqvist,barbieri,dolgov,notzold,hoboya} for large
mixing angle. \emph{Assuming} that the initial momentum spread of
the wave-packet is determined by the large angle scattering mean
free path of charged leptons in the medium\cite{book1}, we find the
following hierarchy between the transverse dispersion
$\tau_{\perp}$, oscillation $T_{med}$, collisional relaxation
$t_{rel}$, Hubble $t_H$,    in-medium coherence $t_{c,m}$ and
longitudinal dispersion $\tau_{\parallel}$ time scales respectively:
for large vacuum mixing angle $\sin2\theta \sim 1$:

\be \tau_{\perp}\ll T_{med} < t_{rel} < t_H \ll t_{c,m} \lesssim
\tau_{\parallel}\ee and for small mixing angle $\sin2\theta \ll 1$

\be \tau_{\perp}\ll t_{rel} \lesssim T_{med} < t_H \ll t_{c,m}
\lesssim \tau_{\parallel}\,.\ee The rapid transverse dispersion is
responsible for the main suppression of both the persistent and
transition probabilities making the amplitudes extremely small on
scales much shorter than any of the other scales. Only a momentum
spread $\sigma \sim \sqrt{|\delta M|^2}$ will make the transverse
dispersion time scale comparable with the oscillation and relaxation
ones. Clearly a better assessment of the momentum spread of
wave-packets in the medium is required to provide a more reliable
estimate of the wave-packet and oscillation dynamics. }

\end{itemize}

\begin{acknowledgments} D.B. thanks  G. Raffelt for illuminating correspondence
and A. Dolgov for discussions. The authors thank L. Wolfenstein for
inspiring conversations and acknowledge support from the US NSF
under grant PHY-0242134. C.M.Ho acknowledges partial support through
the Andrew Mellon Foundation.
\end{acknowledgments}

\end{document}